\documentclass[review]{elsarticle}
\usepackage{amssymb,amsmath}
\usepackage{graphicx}
\usepackage{amsmath}
\usepackage{amssymb}
\usepackage{amsmath}
\usepackage{color}
\usepackage{cleveref}
\journal{Journal of \LaTeX\ Templates}
\usepackage{setspace}
\newcommand{\Z}{\mathbb{Z}}
\newcommand{\N}{{\mathbb{N}}}
\newcommand{\R}{{\mathbb{R}}}

\newcommand{\dis}{\displaystyle}

\usepackage{geometry}

\newgeometry{vmargin={18mm}, hmargin={14mm,18mm}} 
\newtheorem{prop}{Proposition}
\newtheorem{remark}{Remark}
\newgeometry{vmargin={18mm}, hmargin={14mm,18mm}} 
\newtheorem{theorem}{Theorem}

\begin{document}

\begin{frontmatter}

%\title{Canard Phenomenon and MMOs and MMBOs}
\title{Non trivial dynamics in the FizHugh-Rinzel model and  non-homogeneous oscillatory-excitable reaction-diffusions systems}
%%%%%%%%% Insert author address here

\tnotetext[]{Corresponding author\\ Email address: benjamin.ambrosio@univ-lehavre.fr}

\author{B. Ambrosio$^{1,2,\star}$,M. A. Aziz-Alaoui$^1$, Argha Mondal$^{2,3}$, Arnab Mondal$^{2}$, Sanjeev K. Sharma$^{2}$,    Ranjit K. Upadhyay$^{2}$}
\address{
	$^{1}$ Normandie Univ, UNIHAVRE, LMAH, FR-CNRS-3335, ISCN,
	76600 Le Havre, France\\
	$^{2}$ The Hudson School of Mathematics, NJ, USA
	$^{3}$  Department of Mathematics and Computing, Indian Institute of Technology (Indian School of Mines), Dhanbad 826004, India \\
 $^{4}$ Department of Mathematics, Sidho-Kanho-Birsha University, Purulia-723104, WB, India
	$^{5}$ Department of Mathematical Sciences, University of Essex, Wivenhoe Park, UK \\
}

\begin{abstract}
In this article, we discuss the dynamics of the 3-dimensional FitzHugh-Rinzel (FHR) model and a class of non-homogeneous FitzHugh-Nagumo (Nh-FHN) Reaction-Diffusion systems. The Nh-FHN models can be used to generate relevant wave-propagation phenomena in Neuroscience context. This gives raise locally to complex dynamics such as canards, Mixed-Mode Oscillations, Hopf-Bifurcations some of which can be observed in the FHR model. 

\end{abstract}


\end{frontmatter}

%\linenumberso
\section{Introduction and motivation}
The final goal of the present article is to shed light on some mechanisms related to the emergence of complex oscillations (mixed mode oscillations --MMOs--, canards, multiple Hopf Bifurcations...) arising in reaction-diffusion (RD) systems in Neuroscience context. Typically, we consider a non-homogeneous FitzHugh-Nagumo  RD type system (Nh-FHN) in which a parameter is space dependent. The space domain, in this context, typically stands for an excitable tissue. At the center of the spacial domain, the cells are assumed to be in an oscillatory state, and elsewhere in the domain, the cells are in an excitable state. The excitability of the cells depend on the value of a parameter related to the external current, that we allow to vary. Within this context, a remarkable observed phenomenon is the following; upon the variation  of  this parameter value,  waves of depolarization may propagate from the center of the domain toward the boundaries  when the cells are enough excitable, or the solution can evolve to a stationary state if the cells are not enough excitable. For a value of the parameter in between, a bifurcations occurs, typically a Hopf bifurcation. Within this range of parameters, for central cells, mixed mode oscillations or other complex dynamics can be observed in time. This scenario was first described in \cite{Amb-2009} and further analyzed and discussed in a few papers such as \cite{Amb-2017,Amb-2020}. Here, we will consider a slightly different general version of this non-homogeneous FHN RD system (Nh-FHN). Our approach in the present paper is  to emphasize the analogy between such a dynamics observed in the spacial Nh-FHN system for  central cells  and the dynamics of the FitzHugh-Rinzel (FHR) ODE system. This constitutes the main streamline of our article. Accordingly, the general plan of the paper is as follows. After a first introductory part, we will provide an original theoretical and numerical analysis of the FHR system in section 2.  We will then discuss the analogy with the Nh-FHN PDE with illustrative numerical simulations and theoretical results in section 3. Before digging into a more detailed analysis, we introduce the equations to be considered, recalling also some necessary background.
\subsection{The FHR system} 
The FHR ODE model was introduced as a model for bursting oscillations, ($i.e.$ alternated phases of oscillations and quiescent states)   in \cite{Rin-1986,Rin-1987}, and previously studied by J. Rinzel and R. FitzHugh in an unpublished paper in 1979. In those fundamental papers the FHR model reads as:

\begin{equation}
\label{eq:FHR-orig} 
\left\{
\begin{array}{rcl}

\displaystyle \epsilon \frac{dv}{dt}  &=&f(v)-w+y+I,\\

\dis \frac{dw}{dt}  &=& \phi(a+v-bw)\\ 

\dis \frac{dy}{dt}  &=&\epsilon (c-v-dy)\\
\end{array}
\right.
\end{equation}
with
\[f(v)=-(1/3)v^3+v  \]

For the following values of parameters, $I=0.3125, a = 0.7, b= 0.8, c = -0.775, d = 1.0, \phi = 0.08,  \epsilon = 0.0001$ the model exhibits nice bursting behavior. The mechanism is strikingly simple and intuitive. The model consists of a FitzHugh-Nagumo (FHN) system, represented by the two first equations and a super slow variable, whose dynamics are given by the third equation. A relevant approach here is to formally consider $y+I$ as a parameter for the two first equations. Then, the dynamics of the two first equations are known from the FHN analysis. But the variable $y$ moves slowly with the variation of the first variable allowing the emergence of the bursting. The description made in the original papers \cite{Rin-1986,Rin-1987} reflects a precise numerical qualitative analysis of the FHR dynamics supported by numerical illustrations. The author suggested that the model would be suitable for a deeper rigorous mathematical analysis, and invited the community to study the model in this direction. This call would receive an important feedback, and the model was indeed studied later in numerous papers, see for example \cite{Des-2022,Izh-2001,Mon-2019,Woj-2015} to cite only but  a few. Here, we will consider a slightly different version of FHR, for which we will provide numerical and theoretical analysis. The system under consideration writes:
\begin{equation}
\label{eq:FHN3v} 
\left\{
\begin{array}{rcl}

\displaystyle \epsilon \frac{du}{dt}  &=&f(u)-v+w+I,\\

\dis \frac{dv}{dt}  &=& u-bv-c\\ 

\dis \frac{dw}{dt}  &=&\epsilon (-u-w)\\
\end{array}
\right.
\end{equation}
   with $f$ a cubic function, $b>0$, $\epsilon$ a small parameter and $I$ a parameter to vary.
For the ODE section, we will provide an analysis with $f$ and $b$ set as
\[f(u)=-(1/3)u^3+u, b=0.8, c=0.  \]
Next, we move forward with a short presentation of the Nh-FHN model.
\subsection{A non-homogeneous FHN model (Nh-FHN)}
The non-homogeneous FHN reaction-diffusion model (Nh-FHN) to be considered here writes,
\begin{equation}
\label{eq:NhFHN} 
\left\{
\begin{array}{rcl}
\vspace{0.5cm}

\displaystyle \epsilon \frac{du}{dt}  &=&f(u)-v+w+I(x)+du_{xx},\\

\vspace{0.5cm}
\dis \frac{dv}{dt}  &=& u-bv-c(x)\\ 
\end{array}
\right.
\end{equation}
on a real bounded interval space domain $(\alpha,\beta)$ with Neumann Boundary Conditions (NBC). The notation $I(x)$ and $c(x)$ are chosen  to emphasize that they are functions of  the space variable, and that $u_{xx} = \frac{\partial^2 u}{\partial x^2}$. Around the center
of the interval $(\alpha,\beta)$, $I(x)$ and $c(x)$ are set to a value for which the diffusion-less ODE underlying system would generate relaxation oscillations. Out of this region, $I$ and $c$ are set to a value for which the diffusion-less ODE would be in a stationary stable but excitable state. We allow those functions to be varied as a function of a parameter; this leads to a bifurcation path from a stationary state to propagation of oscillations for Nh-FHN. Previously, various studies have been  conducted by some of the authors of the present paper, see \cite{Amb-2009,Amb-2017,Amb-2020}. In section 3, we will present numerical simulations of this PDE to be compared to the dynamics of equation \cref{eq:FHN3v}.

\section{Analysis of the FHR system}
\subsection{A short background on FHN}\label{section1b}
The mechanisms at play in the FHR and Nh-FHN models rely primarily on the dynamics of the FHN model. Consequently, it is worth to briefly first describe the  dynamics of the following classical FHN system

\begin{equation}
\label{eq:FHN}
\left\{
\begin{array}{rcl}
\epsilon \frac{du}{dt}& =& f(u)-v+I\\
	\frac{dv}{dt}& =& u-bv, 
\end{array}
\right.	
\end{equation}
where $f(u)=-\frac{u^3}{3}+u$, $\epsilon$ is a small parameter, $0<b<1$ an $I>0$.
\begin{prop}
\label{prop:FHN2d}
 \Cref{eq:FHN} admits a unique stationary solution $(u^*,v^*)$ given by
\[v^*=u^*/b \]
and $u^*$ is the unique solution of
\[f(u)-u/b+I=0.\]
It follows that $u^*$ is an increasing function of $I$; and the map $I \rightarrow u^*$ is a bijection from $(0,+\infty)$ to $(0,+\infty)$.
Furthermore, as $I$ increases from $0$ to $+\infty$, $(u^*,v^*)$ is successively a repulsive node, a repulsive focus, an attractive focus and an attractive node.\\
At $f'(u^*)=b\epsilon \,\,\,(i.e.\,\, u^*=\sqrt{1-b\epsilon})$ a Hopf bifurcation occurs.
\end{prop}
Furthermore, when $u^*$ is close to zero, the system is known to exhibit relaxation oscillations.
With this result in mind, one principle at play to obtain MMOs becomes clear: if we add a third variable that moves slowly, and in such a way that the dynamics for the system of the two first equations follow the loop: attractive focus, repulsive focus, relaxation oscillation and then return mechanism, MMOs should appear. The next subsection illustrates more in detail this idea to obtain MMOs as well as other phenomena related to the so called canard solutions.
\subsection{A system with MMOs}
We consider here the \Cref{eq:FHN3v} introduced above
\begin{equation*}
\left\{
\begin{array}{rcl}
\epsilon \frac{du}{dt}  &=&f(u)-v+w+I,\\
\frac{dv}{dt}  &=& u-bv\\
\frac{dw}{dt}  &=&\epsilon (-u-w)\\
\end{array}
\right.
\tag{\ref{eq:FHN3v}}
\end{equation*}
with
\[f(u)=-(1/3)u^3+u,  \epsilon=0.1, \, I\in (1.3,1.5), b=0.8. \]
We focus on this system because it exhibits the MMOs.
We describe here two distinct mechanisms leading to alternance of small and large oscillations. The first mechanism corresponds to small oscillations related to the focus nature of the fixed point in the 2-dimensional FHN equation \eqref{eq:FHN} as described in section 1. The second mechanism is different; in this case, the trajectories clearly exhibits canard-type solutions, and the trajectories jump from the repulsive manifold toward alternatively one side or the other side of the stable manifold. Of note, his distinctive evolution toward the left or right side of the stable manifold is made within a tiny region of the repulsive manifold.  We end the section with a remark on Shilnikov Chaos.
\subsubsection{MMOs and focus}
\begin{figure}
	\centering
	\includegraphics[width=5cm,height=5cm]{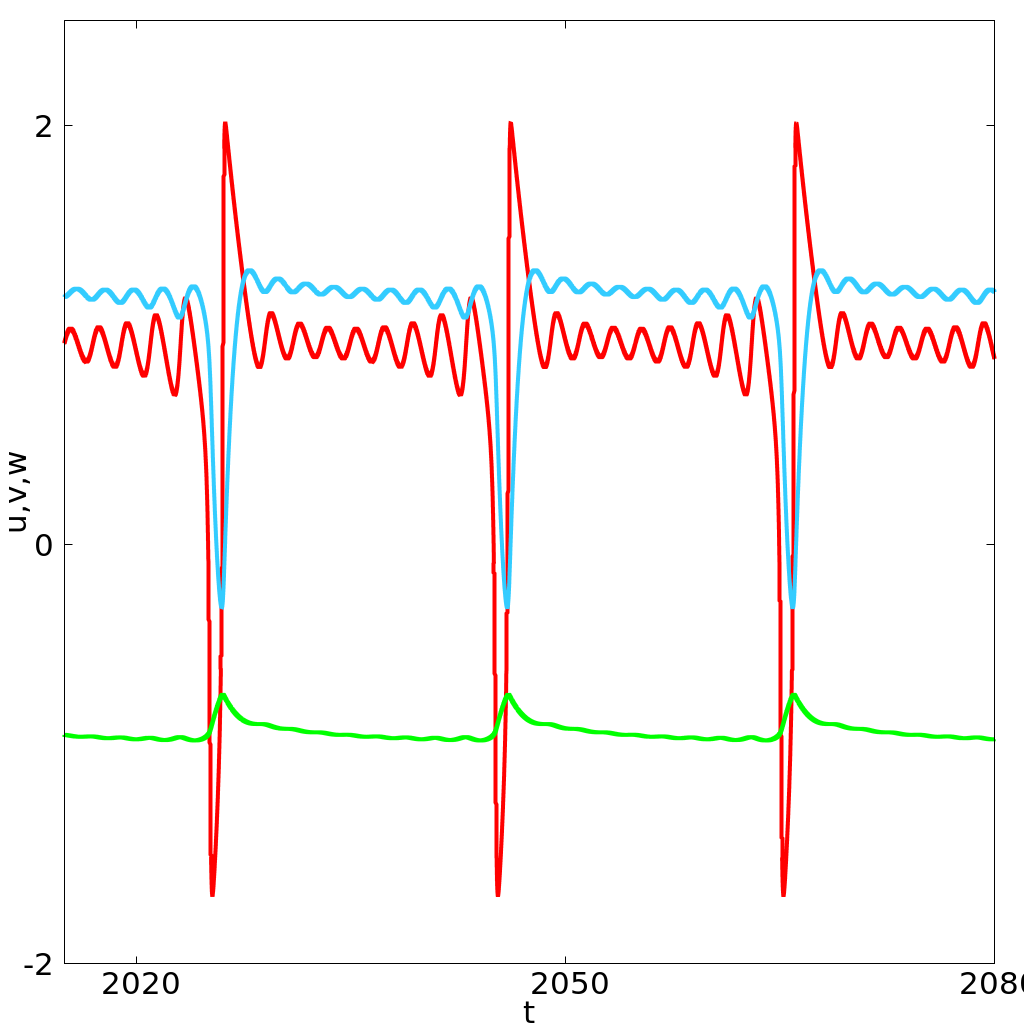}
	\includegraphics[width=5cm,height=5cm]{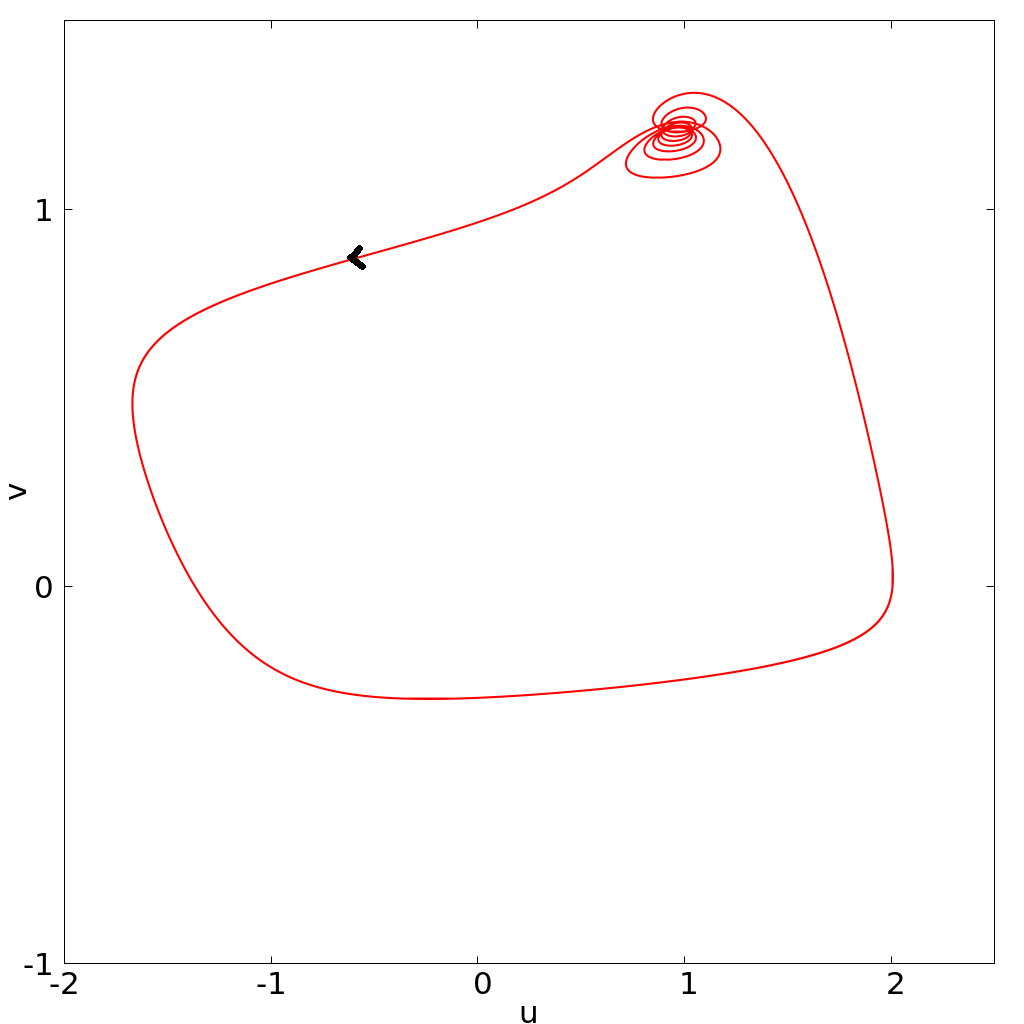}
	\includegraphics[width=5cm,height=5cm]{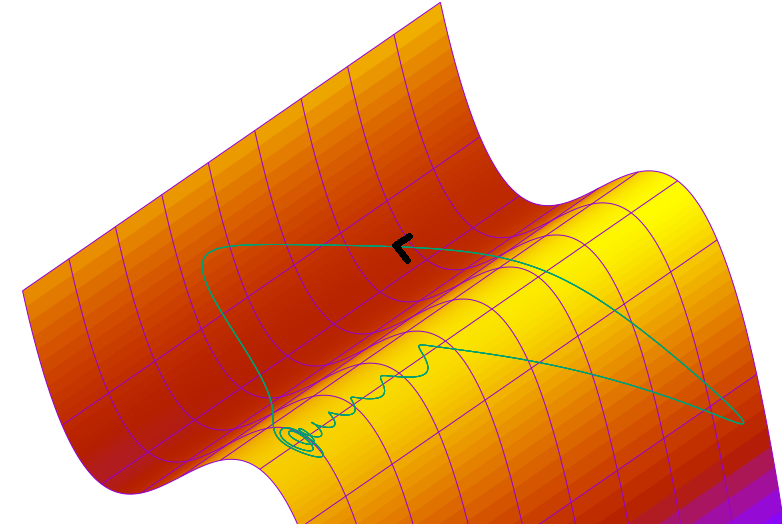}
	\caption{Mixed Mode Oscillations. This figure illustrates the asymptotic behavior of solutions of system \Cref{eq:FHN3v} for $I=1.45$. In the left panel, time series of $u,v$ and $w$ are presented. In the center, we illustrate the projection of the same trajectory on the plane $(u,v)$. In the right, we illustrate the trajectory in the three dimensional phase along with the  critical manifold $v=f(u)+w+I$ . This figure illustrates classical MMOs related to the focus nature of the fixed point of the first two equations (considering $w$ as a fixed parameter). In this case the observed dynamics suggest the following successive states: attractive focus, repulsive focus, relaxation oscillations and return mechanism.} 
	\label{fig:1}
\end{figure}

The first mechanism to obtain MMOs is as follows. Considering $w$ as a parameter, the two first equations represent a classical FHN system, which goes trough a Hopf bifurcation as $w$ is varied. If $w$ varies very slowly, in an interval corresponding to a focus for the FHN system, then we will observe focus-like dynamics in a neighborhood of the fixed point of the FHN subsystem. The focus is first attractive, then repulsive, until the trajectory falls into a relaxation oscillation type. This corresponds to small oscillations followed by a large oscillation. The dynamics are such that during the relaxation oscillation $w$ returns close to its initial value. We have by this means a return mechanism, which starts a new  cycle. This behavior is illustrated in \Cref{fig:1}. 
\begin{remark}
Another approach to describe the above dynamics relies on canards. The small oscillations are also to be seen as trajectories following successively the attractive and repulsive parts of the manifold till they exit the vicinity of the fold trough a relaxation oscillation. We wanted here to emphasize an approach by a dynamical Hopf bifurcation, which appears o be relevant. Of note, in the simulation of \Cref{fig:1}, the trajectory escapes the fold line to the right trough a remarkable long canard type solution. This mechanism will be emphasized in the next paragraph.
\end{remark}
\subsubsection{MMOs and canards}
The second mechanism we want to discuss gives raise to a quite different type of dynamics. In this case, there are no multiple small focus-like oscillations.  Instead, after a  relaxation oscillation, the trajectory may enter a canard type trajectory after crossing the apex of the right part of the critical manifold; it follows the unstable manifold and leaves it either to the left side- where it reaches the left part of the attractive critical manifold- or to the right side - and reaches the right part of the attractive manifold. The latter case corresponds to a small (or middle)  oscillation while the first case gives a large relaxation oscillation. This is illustrated in \Cref{fig:3}. It is worth to describe more in detail the trajectory represented in this figure.  Two large oscillations are followed by a single small oscillation where the trajectory follows the repulsive manifold before being attracted by the right side of the attractive manifold. This is repeated four times, the fifth time, the second large oscillation is replaced by a small oscillation (but larger than the other small oscillations). This corresponds to a brutal change of direction in the trajectory occurring in a tiny zone of the phase space. After that, the cycle is repeated. Introducing the letter M for the medium oscillation, L for large and S for small, we could denote this occurrence: LLS-LLS-LLS-LLS-LM. Note that during successive cycles LLS, the canards occurring during the second L, are significantly different in the phase space, and move forward in a specific direction. At the forth time the canard exits the unstable manifold to the right. Of note, is that this discrimination between left and right exits has been used to generate rich dynamics in various contexts, see the recent papers \cite{Des-2022,Amb-2021}
\begin{remark}
In the next section, computations will show that the regimes of MMOs exhibited below, correspond to a range of parameter value $I$ for which the fixed point has two complex eigenvalues with positive real part and one negative real value. This is a signature of Shilnikov Chaos which relates to dynamics alternating phases on the attractive manifold and phases on the repulsive manifold supported by the complex eigenvalues \cite{Glen-1984,Shi-2007} and references therein.  Figure 2 gives an interesting glimpse. For the initial condition considered here, at the beginning, the solution follows the stable manifold corresponding to the negative eigenvalue. But afterwards, the trajectory exits the neighborhood of the fixed point with a focus-like dynamics; this corresponds to the two complex eigenvalues with positive real parts. After that, the trajectory never comes back in a neighborhood of the fixed point-- the trajectory following the stable manifold corresponds to  a transient behavior-- and as such the asymptotic observed dynamics do not relate to Shilnikov chaos despite the eigenvalues.

\end{remark}

 \begin{figure}
	\centering
	\includegraphics[width=5cm,height=5cm]{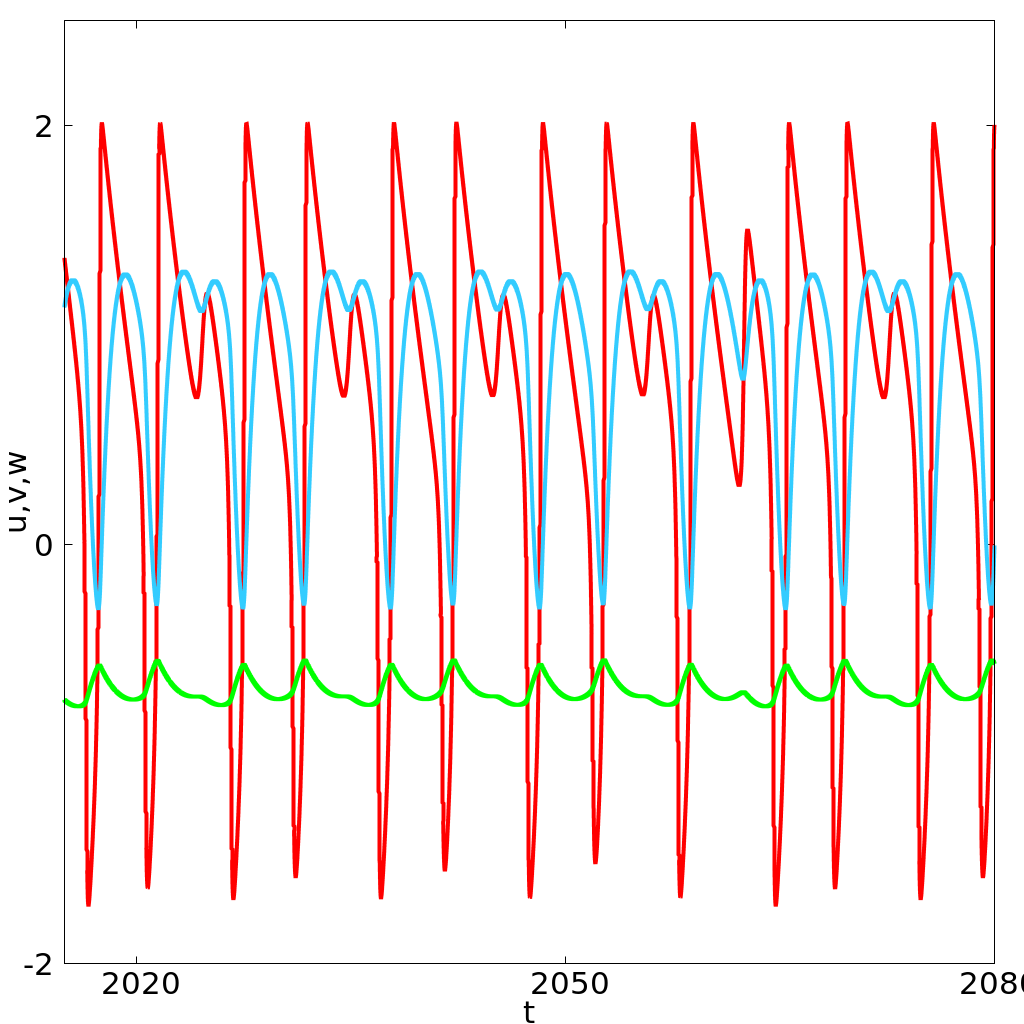}
	\includegraphics[width=5cm,height=5cm]{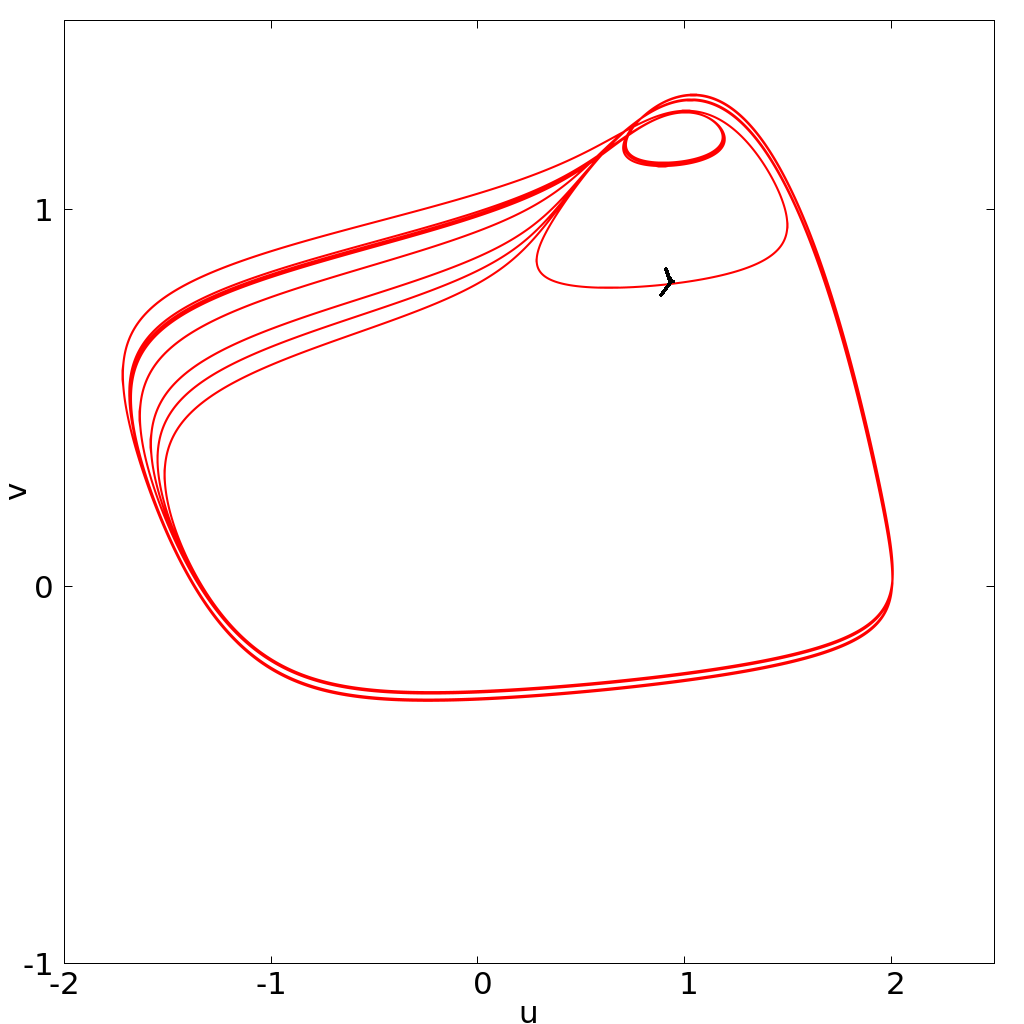}\\
	\includegraphics[width=5cm,height=4cm]{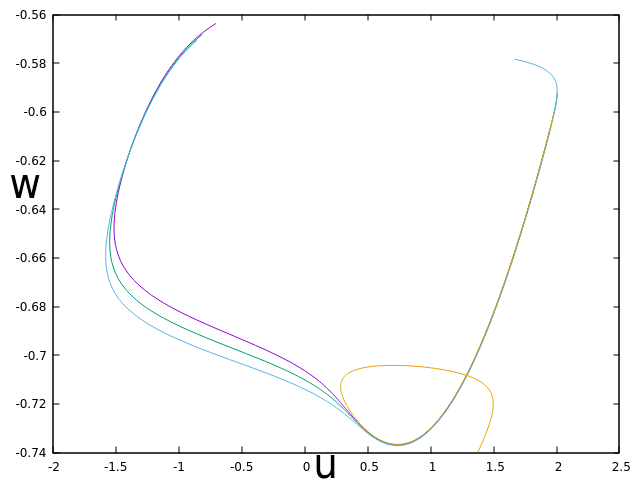}
		\includegraphics[width=5cm,height=4cm]{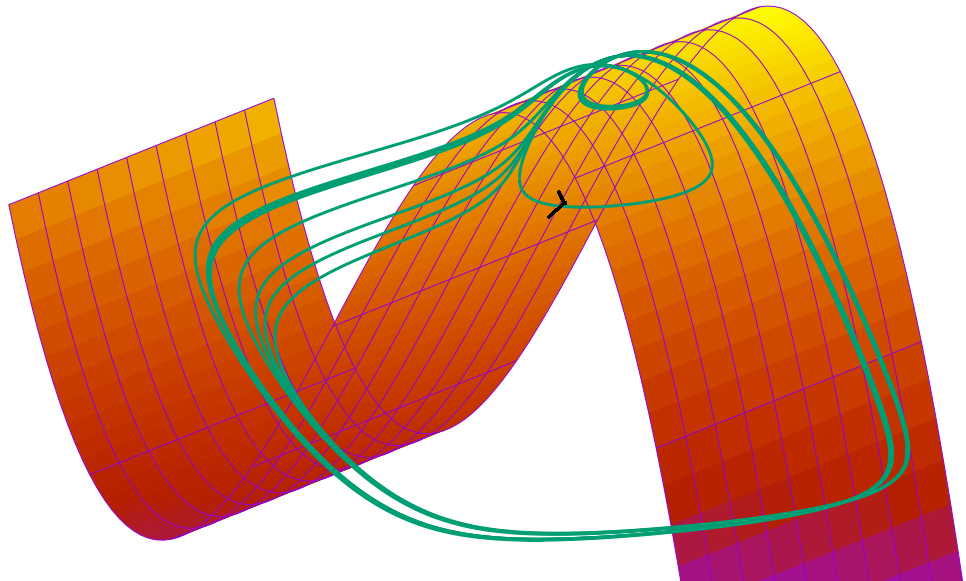}
	
	\caption{(Complex) Mixed Mode Oscillations. In this case, there are no multiple small focus-like oscillations.  Instead, after a  relaxation oscillation, the trajectory may enter a canard type trajectory after crossing the apex of the right part of the critical manifold; it follows the unstable manifold and leaves it either to the left side- where it reaches the left part of the attractive critical manifold- or to the right side - and reaches the right par of the attractive manifold. The latter case corresponds to a small (in fact a middle) oscillation while the first case gives a large relaxation oscillation.  It is worth to describe more in detail the trajectory represented in this figure.  Two large oscillations are followed by a single small oscillation where the trajectory follows the repulsive manifold before being attracted by the right side of the attractive manifold. This is repeated four times, the fifth time, the second large oscillation is replaced by a middle size oscillation. This corresponds to a change in the trajectory occurring in a tiny value range (see bottom left). After that, the cycle is repeated. Introducing the letter M for the medium oscillation, L for large and S for small, we could denote this occurrence: LLSLLSLLSLLSLM. Top left panel: time series of $u$,$v$ and $w$. Top right panel: projection on the $u-v$ plane. Bottom left panel: projection on the $u-w$ plane of four successive segments of the trajectory along the repulsive branch and the subsequent fast trajectory.The first three segment exit towards the left part of the attractive manifold. The forth one exits toward the other side. } 
	\label{fig:2}
\end{figure}

\begin{figure}
	\centering

	\includegraphics[width=7cm,height=7cm]{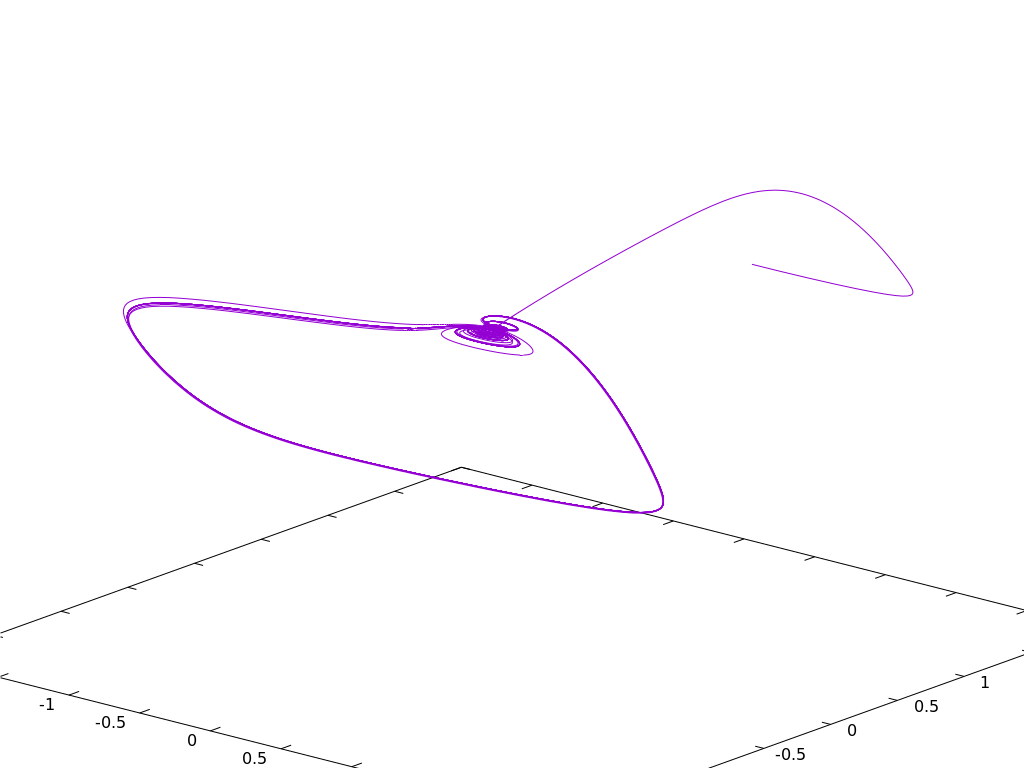}
		\includegraphics[width=7cm,height=7cm]{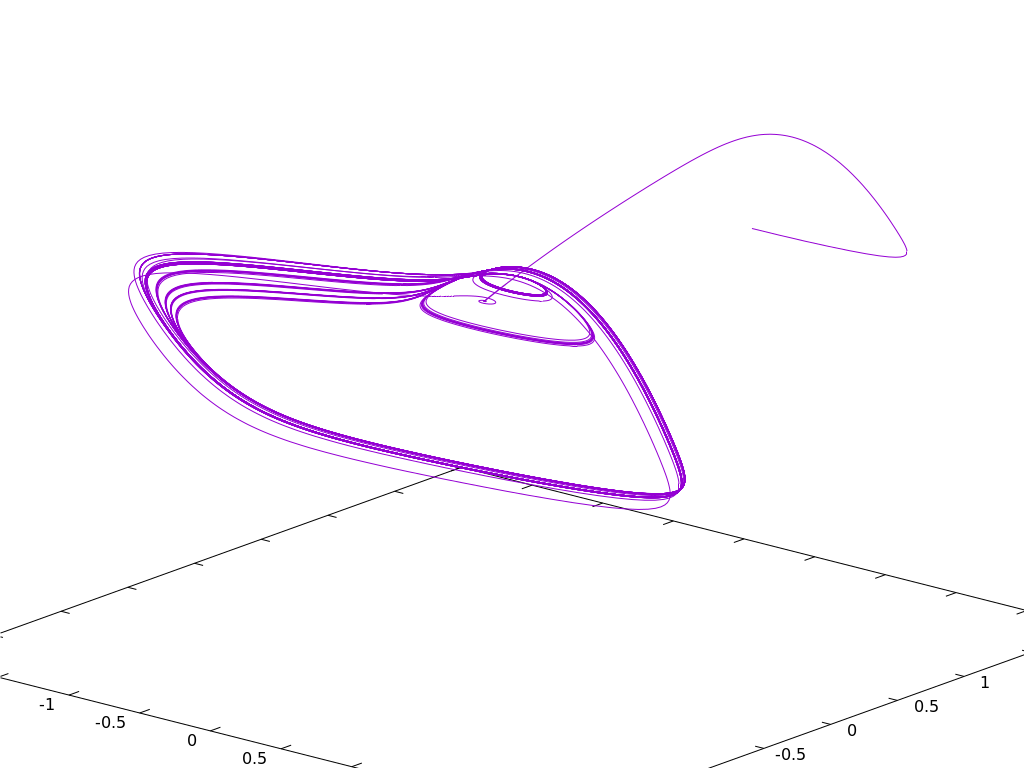}
	
	\caption{This figure is intended to illustrate a remarkable transient behavior. For the parameter values considered here, the stationary point has two complex eigenvalue with positive real parts and one negative real eigenvalue. Accordingly, for the initial conditions chosen here, the trajectory follows the stable manifold associated with negative eigenvalue. After that the trajectories are repelled from the fixed point. The asymptotic behavior, however do not relate so much with those eigenvalues. As such, the Shilnikov chaos is not relevant for thse parameters. Left panel: I=1.45. Right panel: I=1.3. } 
	\label{fig:3}
\end{figure}
\subsection{Basic Stability Analysis}\label{section3}
\begin{figure}
	\centering
\includegraphics[width=5cm,height=5cm]{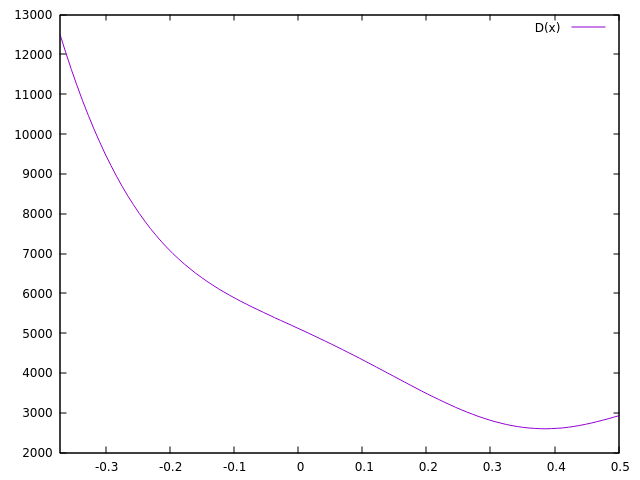}
	\includegraphics[width=5cm,height=5cm]{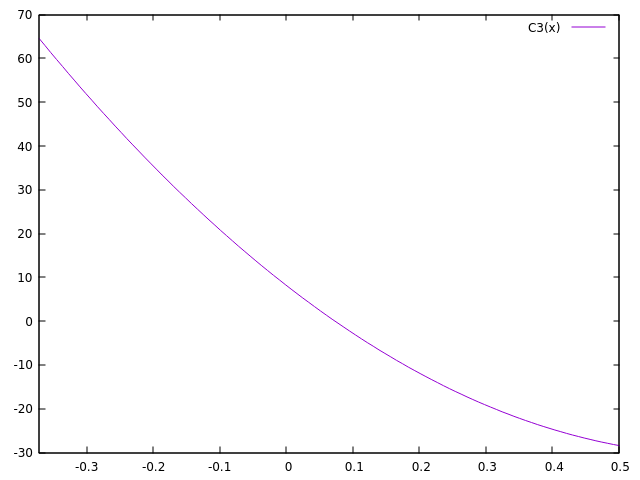}
	\caption{ This figure gives an illustration of \Cref{prop:hopf}. In both panels the $x-$axis represents $f'(u^*)$ for $I\in [1,2]$. Left: illustration of $-\Delta$ as defined in the Cardan formula. Since $\Delta$ is negative, the determinant admits two complex eigenvalues. Right: illustration of $a_2a_1-a_0$ as defined in the Routh-Hurwitz criterion. It is positive for $I>I^*$.  We have also $a_0>0$ and $a_2=\epsilon+b-\frac{f'(u)}{\epsilon}$ which is positive for $I>I^*$, ($I^*$ is the first value for which $a_2a_1-a_0$ or $a_2$ equals zero).  } 
	\label{fig:hopf}
\end{figure}

The following proposition results from simple computations.
\begin{prop}
For any  $b>0$, Eq. \ref{eq:FHN3v} admits a unique fixed point $(u^*,v^*,w^*)$ given by 
\[v^*=u^*/b,\,\,w^*=-u^* \]
where $u^*$ is the unique solution of
\[f(u)-(1+\frac{1}{b})u+I=0.\]
\end{prop}
A local stability analysis provides the following proposition
\begin{prop}
\label{prop:hopf}
 There exists $I^*\in (1,2)$ such that at $I=I^*$, a Hopf bifurcation occurs.
\end{prop}
\textbf{Proof}\\
The jacobian matrix $J(u)$ writes:
\begin{equation}
\begin{pmatrix}
\frac{f'(u)}{\epsilon} &-\frac{1}{\epsilon}&\frac{1}{\epsilon} \\
1&-b&0\\
-\epsilon&0&-\epsilon\\
\end{pmatrix}
\end{equation}
which gives 
\[Det(J(u)-\lambda I)=-\lambda^3+\lambda^2(-\epsilon-b+\frac{f'(u)}{\epsilon})+\lambda(f'(u)+b\frac{f'(u)}{\epsilon}-b\epsilon-\frac{1}{\epsilon}-1)+bf'(u)-1-b\]
\[=-\lambda^3+\lambda^2(-\epsilon-b+\frac{f'(u)}{\epsilon})+\lambda(f'(u)+b\frac{f'(u)}{\epsilon}-b\epsilon-\frac{1}{\epsilon}-1)-bu^2-1\]
Thanks to simple algebraic computations (Routh-Hurwitz criterion and  Cardan formula). One can prove that in the interval $(1,2)$ an eigenvalue is real negative and the two other are complex conjugate; Moreover,
\begin{enumerate}
\item for $I<I^*$ the stationary point is unstable
\item for $I>I^*$, it is stable.
\end{enumerate}
 See \Cref{fig:hopf} for numerical illustration.
\qed

\subsection{Absorbing set and existence of periodic solution, numerical approximation} 
We assume $\epsilon>0$ small enough.
\begin{prop}
System \eqref{eq:FHN3v} admits an absorbing bounded set.

\end{prop}
\textbf{Proof}\\
We define 
\[\psi(t)=\epsilon u^2+v^2+w^2\]
We have
\[\psi'=-\frac{u^4}{3}+u^2+(1-\epsilon)uw+Iu-bv^2-\epsilon w^2\]
By using young inequalities, we can prove that there exist two positive constants $K_1$ and $K_2$ such that

\[\psi'\leq -K_1g+K_2 \]
Multiplying both sides by $e^{K_1t}$ and integrating leads to
\[\psi(t)\leq e^{-K_1t}g_0+\frac{K_2}{K_1}\big( e^{K_1t}-1\big). \]
This completes the proof.\qed

\begin{theorem}
There exists $I_0$, such that for $I \in [0,I_0)$, the system admits a non constant periodic solution. 
\end{theorem}
\textbf{Proof}\\
We assume $I=0$. The result extends to $[0,I_0)$ by continuity.  We consider the Poincar\'e map $F=(F_1,F_2,F_3)$ from the manifold $\mathcal{M}=\{(u,v,w); u=0,v<0, w \in \R\}$ to itself defined thanks to the flow of the ODE. For any $w$ such that $|w|$ not too large, and for $\epsilon$ small enough the map is well defined thanks to the slow-fast theory since the trajectories are close to relaxation oscillations and  $w'$ is of $O(\epsilon)$. We know also that the trajectories will return at  $\mathcal{M}$ with a $v-$coordinate at $f(-1)+O(\epsilon)$ (with $f$ defined in eq. \eqref{eq:FHN3v}, and since $f'(-1)=0$, see classical works on slow-fast systems such as for example \cite{szmolyan2001canards}). From the Brouwer theorem, we deduce that for each fixed $w$, there exists $v$ such that 
\[F_2(0,v,w)=v\]
$i.e$ the $v$-coordinate is a fixed point. By this way, we define a continuous function $w\rightarrow \varphi(w)=v$ such that
\[F_2(0,\varphi(w),w)=\varphi(w).\]
Next, we look for $w^*$ such that,
\[F_3(0,\varphi(w^*),w^*)=w^*.\]

Let's start with an initial condition $w_0<0$ close to $0$. We have
\[F_3(0,\varphi(w_0),w_0)=w_0e^{-\epsilon T}-\epsilon\int_0^Te^{-\epsilon(T-s)}u(s)ds\]
where $T$ is the time at which the flow ensued from $(0,\varphi(w_0),w_0)$ returns to $\mathcal{M}$. Replacing the exponential by its order 1 Taylor expansion, we got that, adopting the notation $F_3(w_0)$ for sake of simplicity
\[F_3(w_0)=w_0-\epsilon w_0T-\epsilon\int_0^Tu(s)ds+o(\epsilon)\]
which gives $F_3(w_0)>w_0$, for $w_0<0$ in a small neighborhood of $0$ and $\epsilon$ small enough. Analog arguments allow to prove that $F_3(w_0)<w_0$ for $w_0>0$ and $\epsilon$ small enough. We omit here the details of the computations which can be made explicit by changing the variable of integration from $s$ to $u$ and rely on the fact that the time spent on the right or left part of the stable manifold depend on the sign of $w$; this corresponds geometrically to the relative position of the nullclines.
 It follows that there exists $w^*$ such that $F_3(w^*)=w^*$ and therefore $(0,\varphi(w^*),w^*)$ is a fixed point of the Poincar\'e map.   
\qed

\subsection{A numerical approximation for small oscillations}
The aim of this section is to illustrate how the small oscillations observed
during typical MMOs (as in figure \Cref{fig:1}) can be locally captured by the dynamics of a ``moving" focus.  
Observation of the small oscillations of \Cref{fig:1} show that oscillations 
in the $(u,v)$ plane are first decreasing in amplitude, then increasing till the trajectory leaves the vicinity of the fold line. Since $w$ moves very slowly, this corresponds to the fact that for fixed $w$ in the corresponding range, the stationary point of the two first equations is a focus, first stable and then unstable, see \Cref{prop:FHN2d}. In this section, we will approximate the dynamics of
the full system by the dynamics of a simpler system for which computations can
be made quite explicitly. We will first operate a change variables to translate our attention on the dynamics around the focus. Then we will look at the linearization and compare the dynamics
of the simple system with the original one.
The next proposition gives the dynamics in a system of coordinates around $(u^*(w),v^*(w))$ where $(u^*(w),v^*(w))$ is the ``stationary'' solution of the subsystem in $(u,v)$ considering $w$ constant. Since $w$ is super slow this system of coordinates is relevant. 
\begin{prop}
Let $(u^*(w),v^*(w))$ be the unique solution of
\begin{equation}
\label{eq:FHN-sta} 
\left\{
\begin{array}{rcl}
0  &=&f(u)-v+w+I,\\
0  &=& u-bv\\
\end{array}
\right.
\end{equation}
Then, after a change of variables around $(u^*(w),v^*(w))$, and with appropriate notations, the trajectory of solutions of \eqref{eq:FHN3v} are given by:
\begin{equation}
\label{eq:FHN-nonauto} 
\left\{
\begin{array}{rcl}
\epsilon \frac{du}{dt}  &=&g(u^*,u)-v +\epsilon^2\frac{1}{f'(u^*)-\frac{1}{b}}(u^*+u+w),\\

\frac{dv}{dt}  &=& u-bv+\epsilon\frac{1}{f'(u^*)-\frac{1}{b}}(u^*+u+w)\\
\frac{dw}{dt}  &=& -\epsilon(w+u^*+u))
\end{array}
\right.
\end{equation}
with
\begin{equation}
\label{eq:g} 
g(u^*,u)=(1-u^{*2})u-u^*u^2-\frac{u^3}{3}
\end{equation}

\end{prop}
Next, our goal is to replace \eqref{eq:FHN-nonauto} by a simpler system that can mimic the small oscillations. 
Now since $w$ is very slow, when $(u,v)$ is small in equation \eqref{eq:FHN-nonauto}, 
it is geometrically relevant to approximate solutions of \eqref{eq:FHN-nonauto}, during the specific period of small oscillations, by the  system:
 \begin{equation}
\label{eq:FHN-nonautolin} 
\left\{
\begin{array}{rcl}
\epsilon \frac{du}{dt}  &=&(1-u^{*2})u-v ,\\

\frac{dv}{dt}  &=& u-bv\\
\frac{dw}{dt}  &=& -\epsilon(w+u^*+u)
\end{array}
\right.
\end{equation}
What is interesting with system \eqref{eq:FHN-nonautolin}, is that it gives you a practical way to compute the number of oscillations involved. Let 
\[A=\begin{pmatrix}
\frac{1}{\epsilon}(1-u^{*2})&-\frac{1}{\epsilon} \\
1&-b\\
\end{pmatrix}\]
And let 
\[\beta=0.5\sqrt{4\det A-(Tr A)^2}\]
The following proposition holds.
\begin{prop}
The number of small oscillations $n_o$ occurring during a time interval $(0,T)$ in \Cref{eq:FHN-nonautolin} is given by
\[n_o=\frac{1}{2\pi}\int_0^T\beta(t)dt\]
\end{prop}
\textbf{Proof}\\
The idea is to work with an approximate solution of \eqref{eq:FHN-nonautolin}. We divide the time interval $(0,T)$ into the subdivision
\[(kdt)_{k\in \{0,1,2,...,N\}},\,\,\mbox{ with }N=\frac{T}{dt}\]

\begin{figure}
 \label{fig:osc-app}
	\centering
	\includegraphics[width=12cm,height=8cm]{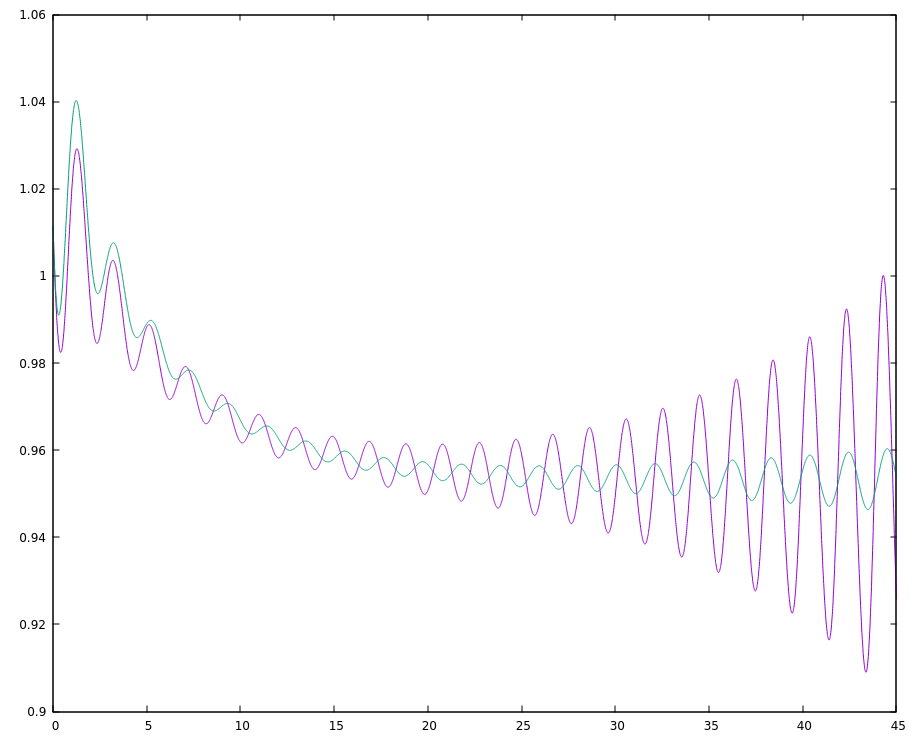}

	\caption{This figure illustrates small oscillations for systems \eqref{eq:FHN-nonautolin}, in green, and \eqref{eq:FHN3v} in purple. At the beginning the solutions are very close, but since we dropped the nonlinear terms the solutions are distinct after some time. Note however that the solutions of \eqref{eq:FHN-nonautolin} capture two aspects of the small oscillations occurring for \eqref{eq:FHN3v}. First, it allows to generate and control in a simple way small oscillations by mimicking the behavior near a focus. Next, it generates oscillations that are first decreasing and then increasing, which corresponds to a dynamical Hopf-Bifurcation behavior ( $Tr A$ changes its sign and $(Tr A)^2-4\det A <0$ ).} 

\end{figure}

Then, at each time step, for fixed $w$, it is possible to compute the solution of the linear subsystem made of the two first equations of \eqref{eq:FHN-nonautolin} on the interval $(kdt,(k+1)dt)$.
Eigenvalues are given by
\[0.5(Tr A_-^+\sqrt{(Tr A)^2-4\det A})\]
and are complex conjugate when
\[1-u^{*2} \simeq b\epsilon\]
Classically, denoting $\alpha+i\beta$ one of this eigenvalues, we have have after a new change of variables,
\[\begin{pmatrix}
\tilde{u}\\
\tilde{v}
\end{pmatrix}
=
\begin{pmatrix}
\alpha +b& \beta \\
1 & 0
\end{pmatrix}
\]
The solution of this last system writes after dropping the tilde
\[u(t)+iv(t)=\exp(\alpha t)\exp(i\beta)\]
Now, the number of small oscillations is given by
\[\sum_k \frac{dt}{2\pi}\beta_k\]
Now, since the approximation converges towards the solution, the results follows from
\[n_0=\lim_{N\rightarrow +\infty}\sum_{k=0}^{N} \frac{dt}{2\pi}\beta_k= \frac{1}{2\pi}\int_0^T\beta(t)dt\]
\qed
\begin{remark}
The interest of this  approximation is that it provides a  geometrical approach to interpret the dynamics of small oscillations occurring in MMOs: the amplitude of the solutions first decrease and then increase. For system \eqref{eq:FHN-nonauto}, this corresponds to $\alpha(u^*(t))<0$, in which case the amplitude decrease and then to $\alpha(u^*(t))>0$ in which case the oscillations increase. 
\end{remark}
\begin{remark}
Also, note that system \eqref{eq:FHN-nonautolin} provides a simple way to generate small oscillations and control the number $n_0$; this approach  captures the essential phenomenon at play in the generation of small oscillations in system \eqref{eq:FHN3v}. \Cref{fig:osc-app} illustrates both oscillations for system \eqref{eq:FHN-nonautolin} and \eqref{eq:FHN3v}. System \eqref{eq:FHN-nonautolin} could be extended to generate this small oscillations recurrently with a reset as it is done in the classical Leaky Integrate and Fire models often used in applications \cite{Abb-1999,Cha-2016,Cha-2018}. The difference being that in LIF models, the equation is  linear and one dimensional with respect to a variable $V$ (the voltage)  with a reset that occurs when a $V$ reaches a threshold value-- while other inputs are typically included to drive this variable $V$. Here, the idea would be to reset the value $w$ when it reaches the desired threshold value. In between the resets, the dynamics are oscillatory in the variables $(u,v)$. 
\end{remark}

\subsection{Slow-Fast Analysis}
In this section, we shall give some insights about the dynamics of \Cref{eq:FHN3v} thanks to a slow-fast approach. Setting $\epsilon=0$ in \Cref{eq:FHN3v} after different time scalings, provides the main dynamical picture. First, setting $\epsilon=0$  gives
\begin{equation}
\left\{
\begin{array}{rcl}
0  &=&f(u)-v+w+I,\\
v_t  &=& u-bv\\
w_t  &=&0\\
\end{array}
\right.
\label{eq:FHN3v-eps0a}
\end{equation}
Then, after the time rescaling $t=\epsilon t'$ in \Cref{eq:FHN3v}, dropping the $'$ and
setting $\epsilon=0$ gives,

\begin{equation}
\left\{
\begin{array}{rcl}
 u_t  &=&f(u)-v+w+I,\\
v_t  &=& 0\\
w_t  &=&0\\
\end{array}
\right.
\label{eq:FHN3v-eps0b}
\end{equation}
Finally,  rescaling  with $t=\frac{1}{\epsilon }t'$ , dropping the $'$ and
setting $\epsilon=0$ gives,
\begin{equation}
\left\{
\begin{array}{rcl}
 0  &=&f(u)-v+w+I,\\
0 &=& u-bv\\
w_t  &=&-(w+u)\\
\end{array}
\right.
\label{eq:FHN3v-eps0c}
\end{equation}
Equations \eqref{eq:FHN3v-eps0a} and \eqref{eq:FHN3v-eps0b} can be seen, for a fixed constant $w$, respectively as the reduced and the layer systems of the classical 2d FHN system. We first consider the fast dynamics. Outside of the critical manifold, the trajectories are given by the layer system \eqref{eq:FHN3v-eps0b}, which is a one dimensional ODE in $u$. For any initial condition (unless we start at the repulsive point or in the stable manifold of a saddle), the trajectory will reach one of the attractive parts of the critical manifold $v=f(u)+w+I$, where $f'(u)<0$. 

Then, we look at the slow dynamics. \Cref{eq:FHN3v-eps0a} is an ODE on the critical manifold.  If the stationary point of \eqref{eq:FHN3v-eps0a} is on the attractive part of the critical manifold,  the solutions are well defined, and thay will evolve to it. To further capture the evolution of the system, we need to consider the very slow motion given by \Cref{eq:FHN3v-eps0c}.  If this stationary point is on the repulsive part with $f'(u)>0$, then when it reaches the fold line $f(u)=0$, system \eqref{eq:FHN3v-eps0a} is not defined, and the derivative explodes in finite time; this corresponds to a jump from one side of the attractive part of the critical manifold to the other one. If the stationary point is on the fold line a more complex behavior (MMOs) can be expected, see \cite{Kru-2001,Kru-2014} and references therein. 
An interesting and less classical insight  from slow-fast analysis relates to  the emergence of the sequence LLSLLSLLSLLSLM described in paragraph 2.2.2. In this case, transition between canards exiting to the left (large oscillation) and canards exiting to the right (middle oscillation) play a crucial role. Following the ideas in  \cite{krupa2008mixedSIAM,de2014three} we highlight hereafter some relevant computations. Consider \Cref{eq:FHN3v}
 \begin{equation*}
 \left\{
 \begin{array}{rcl}
\epsilon u_t &=&f(u)-v+w+I\\
v_t &=& u-bv\\
 w_t  &=&-\epsilon(u+w)
 \end{array}
 \right.
 \end{equation*}
 
 We use the change of variables:
 \[u=1+\bar{u}, v=\tilde{v}+\bar{v}, w=\tilde{w}+\bar{w}\]
 with
  \[f(1)-\tilde{v}+\tilde{w}+I=0\]
  \[1-b\tilde{v}=0\]
  which leads to the equation (dropping the bars)
   \begin{equation*}
 \left\{
 \begin{array}{rcl}
\epsilon u_t &=&-3u^2-u^3-v+w,\\
v_t  &=& u-bv\\
 w_t  &=&-\epsilon(\mu+u+w)
 \end{array}
 \right.
 \end{equation*}
 with
 \[\mu=1+\tilde{w}=-1+\frac{1}{b}-I\]
 Next, we proceed to the change of variables
 \[u=\sqrt{\epsilon}\bar{u}, v=\epsilon\bar{v}, w=\epsilon\bar{w}\]
 Further applying the change of time 
 \[t=\frac{\bar{t}}{\sqrt{\epsilon}}\]
 and dropping the bars, we obtain:
  \begin{equation*}
 \left\{
 \begin{array}{rcl}
 u_t &=&-3u^2-\sqrt{\epsilon}u^3-v+w,\\
v_t  &=& u-b\sqrt{\epsilon}v\\
 w_t  &=&-\sqrt{\epsilon}(\mu+\sqrt{\epsilon}u+\epsilon w)
 \end{array}
 \right.
 \end{equation*}
 Setting $\epsilon=0$ gives:
   \begin{equation}
 \left\{
 \begin{array}{rcl}
 u_t &=&-3u^2-v+w,\\
\frac{dv}{dt}  &=& u\\
 \frac{dw}{dt}  &=&0
 \end{array}
 \right.
 \label{eq:integrable}
 \end{equation}

 \begin{prop}
The point $(u,v)=(0,w)$ is a stationary point of center type for the projection of system \eqref{eq:integrable} into the $(u,v)$-plane. Every trajectory passing trough $(0,c)$ with $w<c<w+\frac{1}{6}$ is a periodic solution. Every solution passing trough $(0,c)$ with $c\geq w+\frac{1}{6}$ satisfies $u(t)\rightarrow -\infty$ and $v(t)\rightarrow -\infty$ as $t\rightarrow t^*$ where $t^*$ is the maximal time for which the solution is defined. Furthermore as $c$ goes to $w+\frac{1}{6}$ the diameter of the periodic solution goes to $+\infty$. 
\end{prop}

\begin{figure}
     \centering
     \includegraphics[width=5cm,height=5cm]{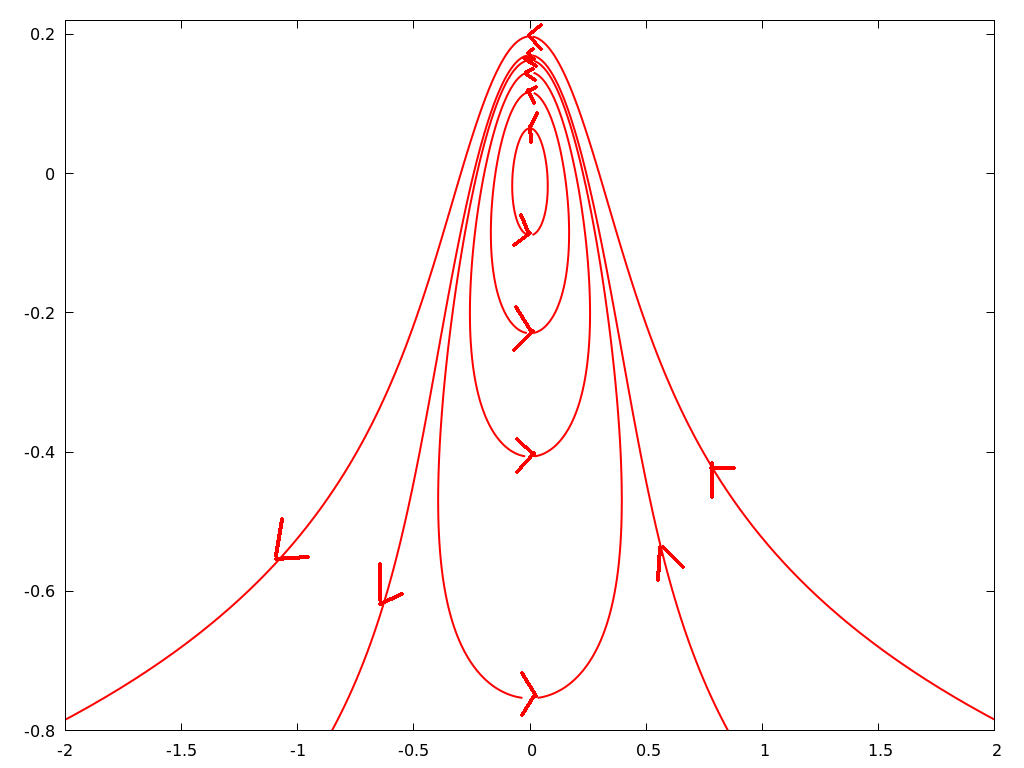}
     \caption{Solutions of the integrable \Cref{eq:integrable}. The fixed point lies in the middle of limit cycles (nonlinear center). Above some treshold, solutions are no longer periodic but rather escape the vicinity of the stationary point. This illustrates the behavior of \Cref{fig:2} : some canards exit to the left while other exit to the right.}
     \label{fig:ls}
 \end{figure}

\textbf{Proof}\\
Without loss of generality we work with $w=0$, since the general case follows from the change of variable
\[-\tilde{v}+w=-v\]
Next, we note that the function
 \[G(u,v)=(-3u^2-v+1/6)\exp(6v).\]
 is a first integral of equation \eqref{eq:integrable} (with $w=0$). It follows that
the solutions of \eqref{eq:integrable} are contained in the level sets of the function $G$. 
The proposition then follows from quite long yet elementary computations to explicit the details of the level sets of $G$. We avoid the details here. An illustrative picture is provided in \Cref{fig:ls} .
 \qed
 
% \begin{equation*}
% \left\{
% \begin{array}{rcl}
% 0 &=&f(u)-v+w+I,\\
% \frac{dv}{dt}  &=& u-bv\\
% \frac{dw}{dt}  &=&0
% \end{array}
% \right.

% \end{equation*}   
% After change of time, $t=\epsilon t'$, and setting $\epsilon=0$ (we drop the prime), leads to the layer system

% \begin{equation*}
% \left\{
% \begin{array}{rcl}
% \frac{du}{dt}  &=&f(u)-v+w+I,\\
% \frac{dv}{dt}  &=& 0\\
% \frac{dw}{dt}  &=&0\\
% \end{array}
% \right.

% \end{equation*}
% Note that the two above equations are those obtained for a two dimensional ODE system. Lastly, we proceed to the change of variable $t=(1/\epsilon)t'$ and set $\epsilon=0$. This gives 

% \begin{equation*}
% \left\{
% \begin{array}{rcl}
% 0 &=&f(u)-v+w+I,\\
% 0 &=&(u-bv) \\
% \frac{dw}{dt}  &=&-(w+u)\\
% \end{array}
% \right.

% \end{equation*}
% which can also be written
% \begin{equation*}
% \left\{
% \begin{array}{rcl}
% 0 &=&f(u)-v+w+I,\\
% 0 &=&(u-bv) \\
% \frac{dw}{dt}  &=&-(w+g^{-1}(w))\\
% \end{array}
% \right.

% \end{equation*}
% where $g^{-1}(w)$ is solution of $f(u)-u/b+w+I=0 $ and corresponds to the stationary solution for each fixed $w$.
% This idea has been used in nulerical simulations. 
% Now for 2d, relaxation oscillations, canards, small oscillations.
% ADD THE QUADRATIC EQUATION.

\section{Dynamics in the NhFHN model}
In this section we shall consider the following system 

\begin{equation}
\label{eq:NhFHN} 
\left\{
\begin{array}{rcl}
\vspace{0.5cm}

\displaystyle \epsilon \frac{du}{dt}  &=&f(u)-v+I(x)+du_{xx},\\

\vspace{0.5cm}
\dis \frac{dv}{dt}  &=& u-bv-c(x)\\ 
\end{array}
\right.
\end{equation}
on a real segment $(\alpha,\beta)$ with Neumann Boundary Conditions (NBC). The notations $I(x)$ and $c(x)$ are to emphasize that these parameters can be made dependent on the space variable.

We first look for stationary solutions of \eqref{eq:NhFHN}. The equation writes:
\begin{equation}
\label{eq:NhFHN-sta1} 
\left\{
\begin{array}{rcl}
\vspace{0.5cm}

\displaystyle 0 &=&f(u)-v+I(x)+du_{xx},\\

\vspace{0.5cm}
\dis 0 &=& u-bv-c(x)\\ 
\end{array}
\right.
\end{equation}

\begin{theorem}
Equation \eqref{eq:NhFHN-sta1} admits a solution
\end{theorem}
\textbf{Proof}
\eqref{eq:NhFHN-sta1} is equivalent to
\begin{equation}
\label{eq:NhFHN-sta2} 
\left\{
\begin{array}{rcl}
\vspace{0.5cm}

\displaystyle 0 &=&f(u)-(u/b)+(1/b)c(x)+I(x)+du_{xx},\\

\vspace{0.5cm}
\dis v &=&(1/b)(u-c(x)) \\ 
\end{array}
\right.
\end{equation}
The proof relies on  the Leray-Shauder degree theory, see appendix.\\
We consider the Banach space $E=L^6(a,b)$ with the usual norm. Let $B(0,r)$ be the ball of radius $r$ of $E$. By definition of the degree $d$, for any $r>0$, 
\[d(Id,B(0,r),0)=1\]
Next, for a given function $\bar{u}$ in $L^6$, we define $F(\bar{u})$ as the solution of
\begin{equation}
\label{eq:NhFHN-sta2-1d} 
\displaystyle -du_{xx}+\bar{u}/b =f(\bar{u})+(1/b)c(x)+I(x)\\
\end{equation}
In the remaining of the proof, without loss of generality we simply replace $(1/b)c(x)+I(x)$ by $I(x)$. Thanks to the theory of the topological degree, to prove the existence of a solution, it is sufficient to prove that for all $t \in [0,1]$ the equation 
\begin{equation}
    u-tF(u)=0 \label{eq:topdegree-t} 
\end{equation}
doesn't admit any solution on the sphere of $L^6$, $i.e$, $\partial B(0,r)$, for a $r$ sufficiently large. This indeed ensures that
\[d(Id-tF,B(0,r),0)=d(Id,B(0,r),0)=1\]
for all $t\in [0,1]$, which for $t=1$ proves the existence of a solution.
By replacing $\bar{u}$ by $u$ and $u_{xx}+(1/b)u$ by $\frac{1}{t}u_{xx}+(1/tb)u$, we find that  this equation is equivalent to
\[\displaystyle -du_{xx}+u/b=tf(u)+tI(x)\]
Multiplying both sides by $u$ and integrating leads to
\[\displaystyle d\int u^2_{x} +(1/b)\int u^2=t\bigg(-\int u^4+3\int u^2+\int uI\bigg)\]
Note that there exists thre positive constants $\alpha, \beta$ and $C$ such that the right hand side is less than
\[-\alpha ||u||^2_{L^2}+\beta+C||u||_{L^2} \]
It follows that
\[||u||_{L^2}\leq C_1 \mbox{ for some constant } C_1\]
and
\[||u_x||_{L^2}\leq C_2 \mbox{ for another constant } C_2\]
Now, from the continuous injection from $H^1$ into $L^6$
\[||u||_{L^6}\leq C_3||u||_{H^1}\leq C_4\]
Now, choosing $R>C_4$, implies that \Cref{eq:topdegree-t} doesn't admit a solution with $||u||_{L^6}=R$.
\qed
\begin{remark}
Note that since we work in 1d, we could also study the resulting ODE
\begin{equation}
\label{eq:NhFHN-sta2-ODE} 
\left\{
\begin{array}{rcl}
\vspace{0.5cm}

\dis u' &=&w \\

\vspace{0.5cm}
\displaystyle dw' &=&-f(u)+(u/b)-I(x),\\

\vspace{0.5cm}
\dis v &=&(1/b)u \\ 
\end{array}
\right.
\end{equation}
\end{remark}
\subsection{Stability analysis}
Let us denote by $(\bar{u},\bar{v})$ a solution of \Cref{eq:FHN-sta}. Equation \eqref{eq:NhFHN} around $(\bar{u},\bar{v})$ rewrites  
\begin{equation}
\label{eq:NhFHN-centre} 
\left\{
\begin{array}{rcl}
\vspace{0.5cm}

\displaystyle \epsilon \frac{du}{dt}  &=&f'(\bar{u})u-v-\bar{u}u^2-\frac{1}{3}u^3+du_{xx},\\

\vspace{0.5cm}
\dis \frac{dv}{dt}  &=& u-bv\\ 
\end{array}
\right.
\end{equation}

The resulting linearized equation writes
\begin{equation}
\label{eq:NhFHN-lin} 
\left\{
\begin{array}{rcl}
\vspace{0.5cm}

\displaystyle \epsilon \frac{du}{dt}  &=&f'(\bar{u})u-v+du_{xx},\\

\vspace{0.5cm}
\dis \frac{dv}{dt}  &=& u-bv\\ 
\end{array}
\right.
\end{equation}

We now proceed as in \cite{Amb-2020}, and are interested in the following equation

\begin{equation}
\label{eq:SL}
f'(\bar{u})u+d u_{xx} =\lambda u
\end{equation}
Equation \eqref{eq:SL} is a regular Sturm-Liouville problem for which the classical following theorem, see for example \cite{BookTeschl}, p 160-162, (see also for example \cite{Gil-1977,Jos-2013}) holds.
\begin{theorem}
\label{th:spec}
There exists an increasing sequence of real numbers $(\lambda_k)$ and an orthogonal basis $(\varphi)_{k\in \N}$ of $L^2(a,b)$ such that:
\begin{equation}
\begin{array}{rcl}
\label{SL}
d\varphi_{kxx}+f'(\bar{u})\varphi_k&=&-\lambda_k \varphi_k\\
\varphi_k'(a)=\varphi'_k(b)&=&0. 
\end{array}
\end{equation}
Furthermore, 
\[\lim_{k \rightarrow +\infty}\lambda_k=+\infty,\]
\begin{equation}
\label{SL}
\lambda_0 = \inf_{u \in H^2(a,b);\int_a^bu^2dx=1} d\int_a^bu_x^2dx-\int_a^b f'(\bar{u})u^2dx,
\end{equation} 
\[\lambda_0\geq -1,\]

and,
\[\lambda_k=\frac{\pi^2k^2}{4a^2}+O(k)\]
(Weyl assymptotics)
\end{theorem}

  The behavior of the linearized system \eqref{eq:NhFHN-lin} is determined by the sequence of linearized two-dimensional equations:
  \begin{equation}
\label{eq:NhFHN-lin} 
(E_k)\left\{
\begin{array}{rcl}
\vspace{0.5cm}

\displaystyle \epsilon \frac{du_k}{dt}  &=&-\lambda_ku_k-v_k,\\

\vspace{0.5cm}
\dis \frac{dv}{dt}  &=& u_k-bv_k\\ 
\end{array}
\right.
\end{equation}
The following result follows from simple computations
  \begin{prop}
  Eigenvalues of $(E_k)$ are given by
  \[\mu^k_{1}=\frac{1}{2\epsilon}\big(-\lambda_k-b\epsilon-\sqrt{(\lambda_k-b\epsilon)^2-4\epsilon(1+b\lambda_k)} \big)\]
  and
  \[\mu^k_{2}=\frac{1}{2\epsilon}\big(-\lambda_k-b\epsilon+\sqrt{(\lambda_k-b\epsilon)^2-4\epsilon(1+b\lambda_k)} \big)\]
 It follows that:
  \[\mu^k_{2}=-3b\epsilon+o(\frac{1}{\lambda_k}) \]

  \end{prop}
The following theorem holds.

\begin{theorem}[Stability in $L^2$]
Assume that $\lambda_0>-b\epsilon$ then for any $\mu>0$, there exists a sequence of positive numbers $(\mu_k)$ such that if for all $k$ $u^2_k(0)+v_k^2(0)<\mu_k$ then the solution $(u,v)$ of \eqref{eq:NhFHN-centre} satisfies
\[ \forall t>0,\,\,||u,v)(t)||_{L^2}<\mu \]

\end{theorem}
\textbf{Proof}\\
The proof relies on the fact that the eigenvalues have a real negative part. For the details of the proof, we refer to \cite{Amb-2020} where a similar result if proved for the case $b=0$.

\qed

\begin{prop}[Instability]
Assume 
\[\int_a^bf'(\bar{u})>0\]
then for $\epsilon$ small enough, $\lambda_0<0$ and the real part of $\mu_2^0$ is positive. 
\end{prop}

The proof of the following proposition is straightforward if one choose $c$ and $I$ not depending on $x$. The interesting cases arise when $c$ and $I$ are depending on $x$. Such examples have been provided in \cite{Amb-2017,Amb-2020}. 
 \begin{prop}
There exists a continuous path $(c_p(x),I_p(x)),\,\, p\in [0,1]$, such that for $p=1$, the stationary solution is stable, and for $p=0$ is unstable. For some $p^*\in (0,1)$, a Hopf bifurcation occurs. 
\end{prop}

\subsection{A few complex oscillations in the NhFHN model} 
\subsubsection{Filtering of frequencies and local MMOs }
\label{parag:MMOSPDE1}
\begin{figure}
	\centering
	\includegraphics[width=5.5cm,height=4.5cm]{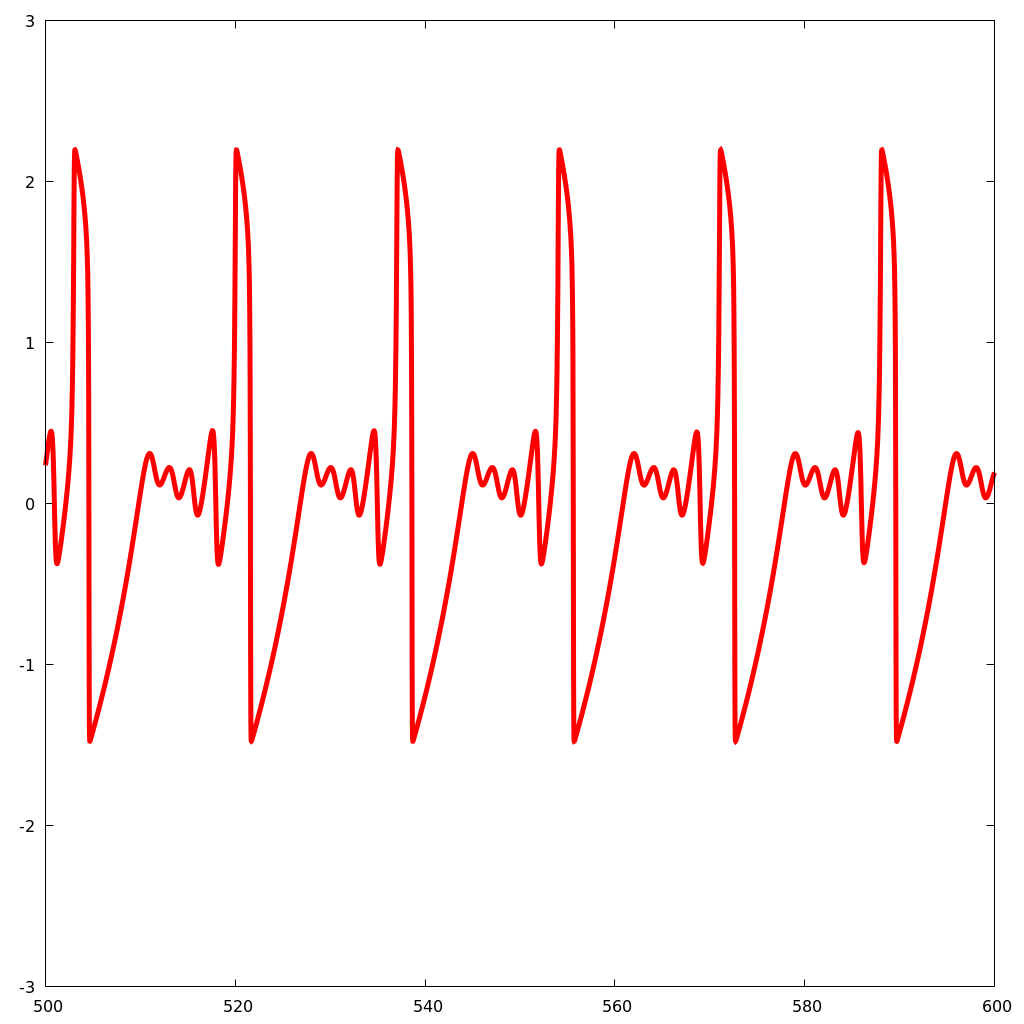}
		\includegraphics[width=5.5cm,height=4.5cm]{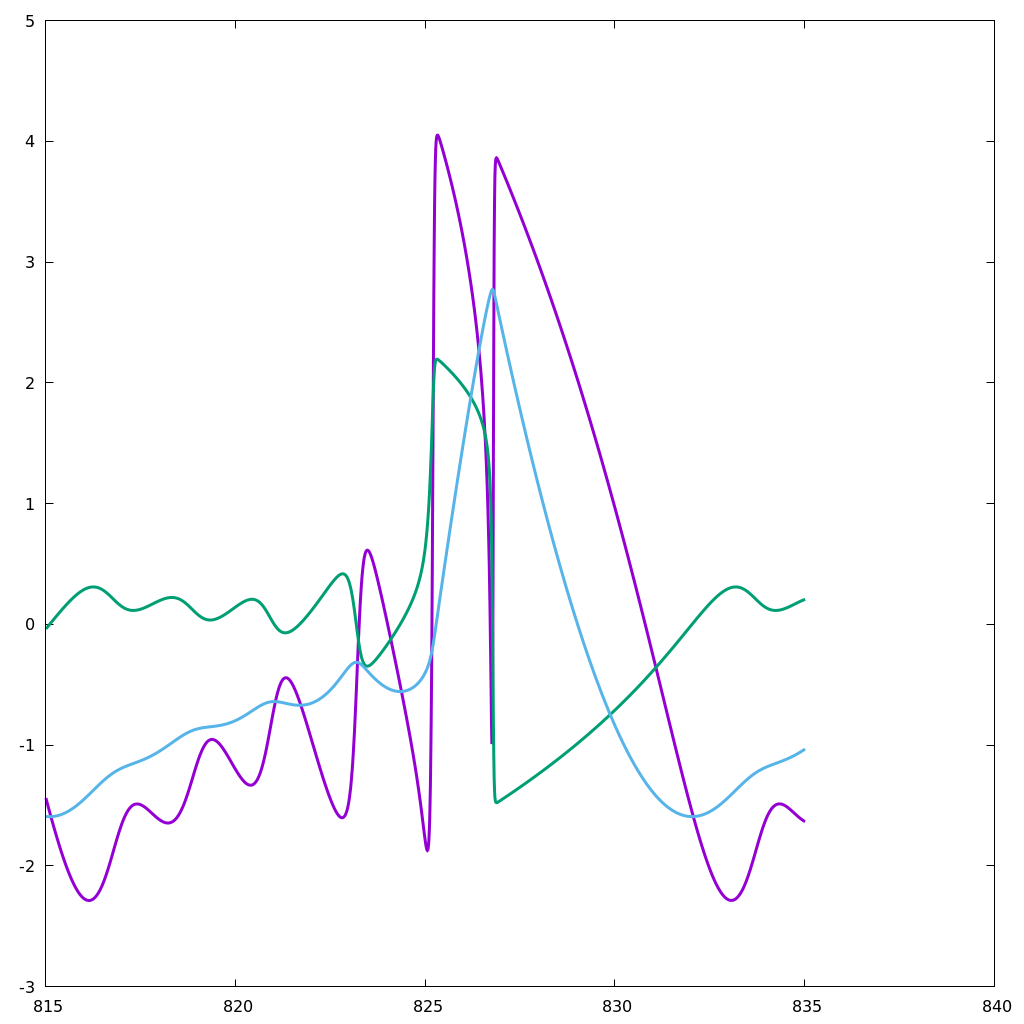}
		\includegraphics[width=5.5cm,height=4.5cm]{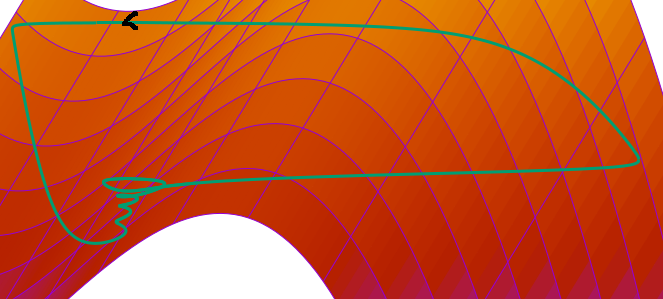}\\[5mm] 
		
			\includegraphics[width=5.5cm,height=4.5cm]{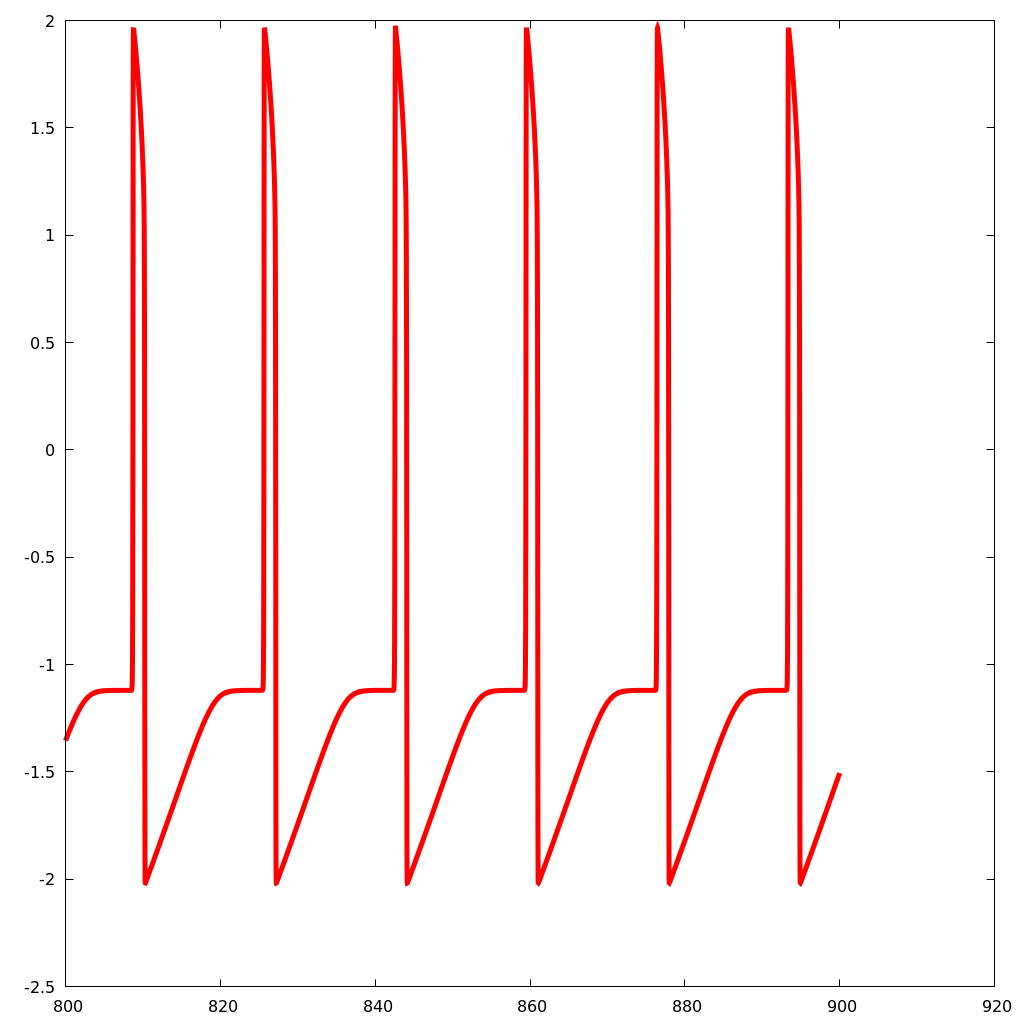}
			\includegraphics[width=5.5cm,height=4.5cm]{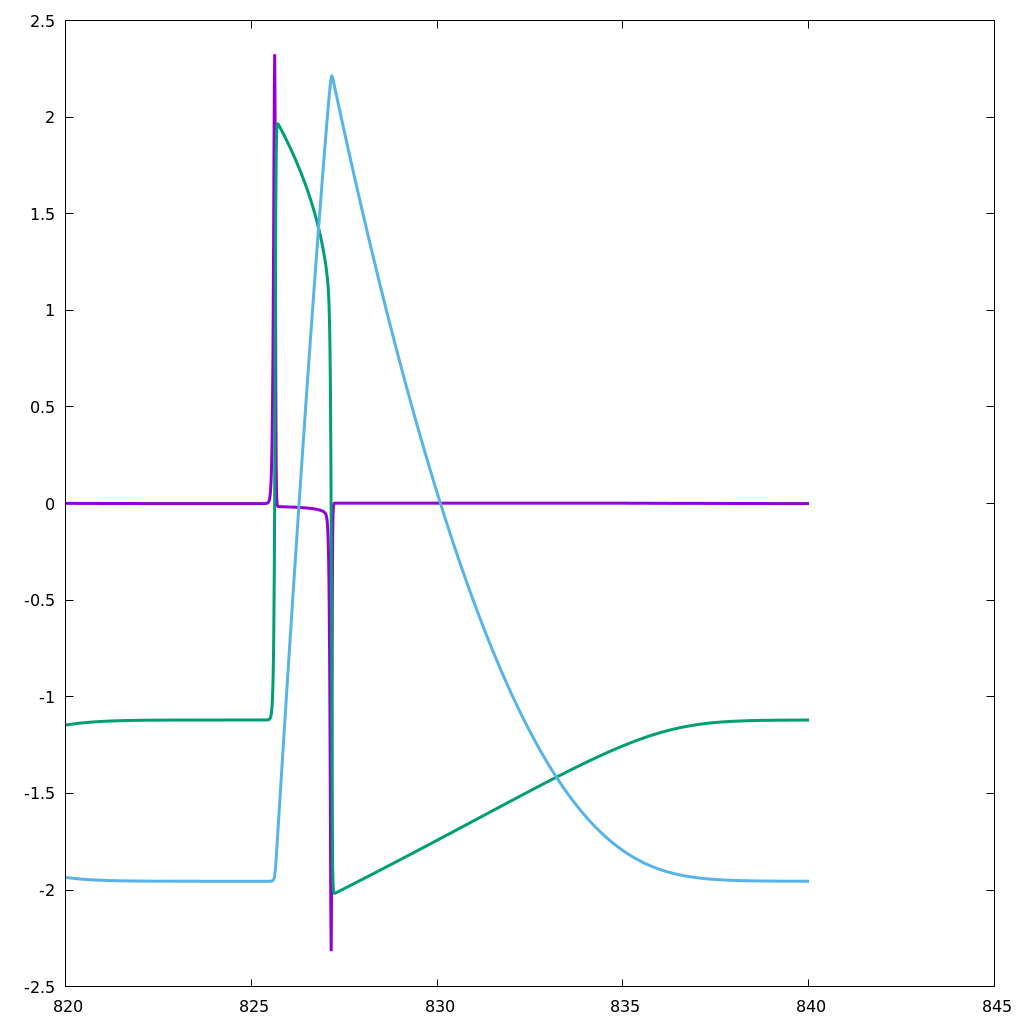}
	\includegraphics[width=5.5cm,height=4.5cm]{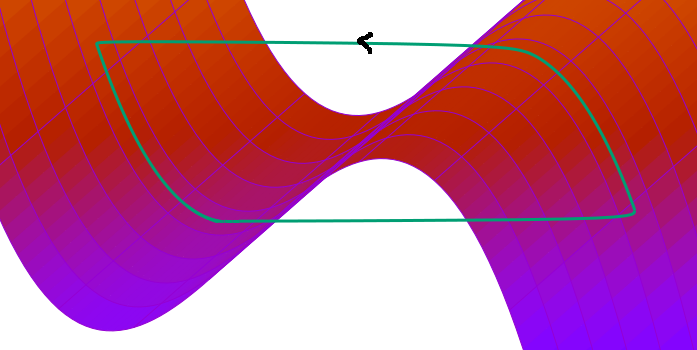}
	\caption{This figures illustrates the simulation of \Cref{eq:NhFHN} with the set of parameters given in \ref{eq:parametersPDE1}. The first row corresponds to a fixed value of $x$ near the center for which $c(x)=0$. The second row of the figure corresponds to value of $x$ near the left border for which $c(x)=c_0=-1.12$. In the first column, we represent $u$ as a function of time. In the second column, we represent $u,v$ and $u_{xx}$ as functions of time respectively in green, blue and violet. In the third column, we represent the trajectory $u,v,u_{xx}$ along with the manifold $v=f(u)+du_{xx}$ (where $u_xx$ is seen as the third variable). This figure illustrates, the propagation of relaxation oscillations from the center cells toward the borders with a filtration of small oscillations observed in the center cells. Note that as a result, the frequency of oscillations is reduced in comparison with the frequency of the diffusionless system (with $c=0$). We refer to \Cref{parag:MMOSPDE1} for more detailed explanation.    }
	\label{fig:PDE1}
\end{figure}

In this paragraph, we shall discuss some qualitative dynamics arising near a Hopf bifurcation. We illustrate the numerical simulation of system
\eqref{eq:NhFHN} with the following set of parameters
\begin{equation}
\label{eq:parametersPDE1}
    \begin{array}{rl}
         f(u)=&-u^3+3u  \\
         c(x)=& 0   \,\,  \mbox{ in the center}\\
         c(x)=&c_0=-1.12   \,\,  \mbox{ otherwise} 
    \end{array}
\end{equation},  
It is known (see \cite{Amb-2009,Amb-2017,Amb-2020}) that if $c_0$ is decreased, the stationary solution of \eqref{eq:NhFHN} is stable, and if it is very close to $-1$ propagation of relaxation oscillations occur. For the present value of $c_0$, we are close to a bifurcation point where more complex phenomenon is observed. Specifically, in this case, we observe mixed mode oscillations for a center cell. Relaxation oscillations will propagate at a frequency smaller than the natural frequency of FHN in the oscillatory regime. This is illustrated in  \Cref{fig:PDE1}. In this figure, the first row corresponds to a fixed value of $x$ near the center.  At this point of the space $c(x)=0$. The second row of the figure corresponds to value of $x$ near the left border. For this latter value of $x$, we have $c(x)=c_0=-1.12$. We observe propagation of oscillations  from the center toward the border. Note however that only large oscillations propagate. Small oscillations occurring in the center cells are filtered. The second row allows us to provide a clear interpretation of the local dynamics related to wave propagation. In the bottom left panel, we represent $u$ for fixed $x$ as a function of time. We observe at this point relaxation oscillations at a lower frequency than the oscillatory $FHN$ ($i.e$ the diffusionless system for $c=0$). In the bottom center panel, we represent $u,v$ and $u_{xx}$ as functions of time, respectively with the colors green, blue and purple. Trough this panel, one can see how the wave propagation is seen locally. It corresponds to a wave of depolarization coming from the right. When, the right neighboring cell jumps, the term $u_{xx}$ becomes positive and jumps too. This causes the system to leave the vicinity of the resting point and in turn to jump toward the right part of the stable manifold (see the bottom right panel). Then the diffusion comes back to almost zero again. The same dynamic occurs, when the cell goes from the right part of the stable manifold towards the left part of the stable manifold, with an excursion of the diffusion in the negative values. The dynamics of $v$ follows simply from the fact that $v_t=u-c$. The bottom right panel illustrates the dynamics of $u,v$ and $u_{xx}$ in the three dimensional space. The ``critical" manifold $v=f(u)+du_{xx}$ is also represented where $u_{xx}$ is considered as a variable.   It illustrates the oscillations of relaxation for the cells near the border. For the first row, the panels are drawn in the same way. For this cell the dynamics resemble the MMOS pictured for the \Cref{eq:FHN3v}, see \Cref{fig:1}.

\subsubsection{Fade of wave propagation (Death spot) }
\label{parag:MMOSPDE2}
In this paragraph, we discuss another phenomenon arising as we vary a parameter (denoted by $p$ in this paragraph). We present here the numerical simulation of system
\eqref{eq:NhFHN} with the following set of parameters
\begin{equation}
\label{eq:parametersPDE2}
    \begin{array}{rl}
         f(u)=&-u^3+3u  \\
         c(x)=& p(x/\beta)^4-2p(x/\beta)^2    \\
         \beta>0&\alpha=-\beta
    \end{array}
\end{equation}  

\begin{figure}
	\centering
	\includegraphics[width=5.5cm,height=4.5cm]{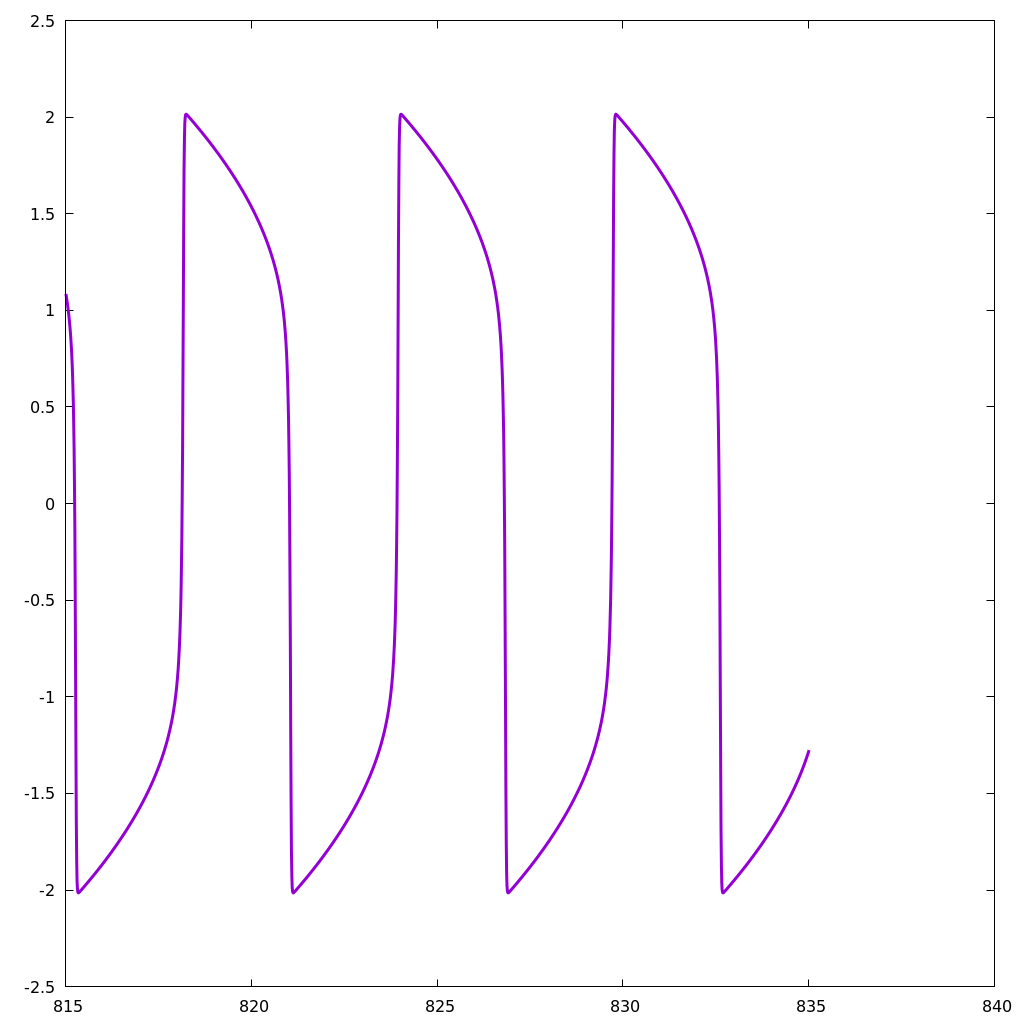}
		\includegraphics[width=5.5cm,height=4.5cm]{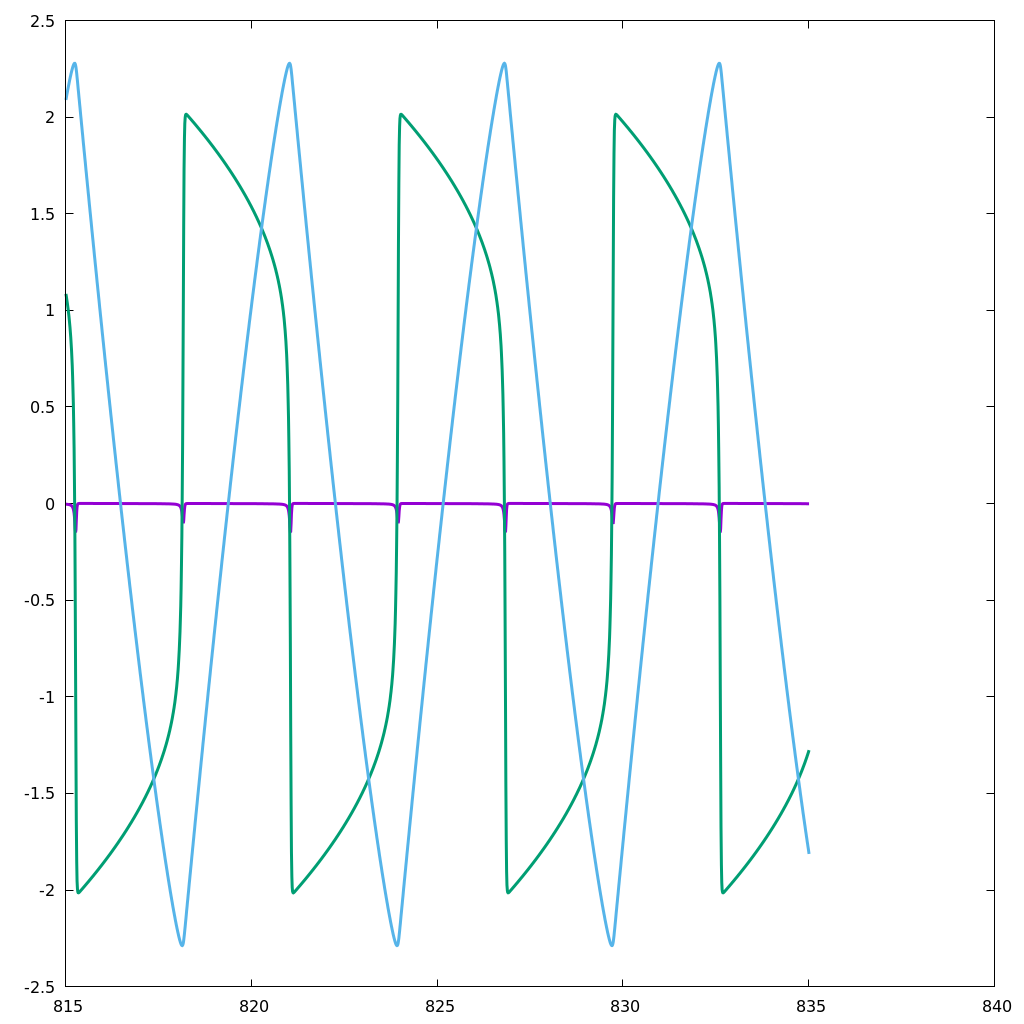}
		\includegraphics[width=5.5cm,height=4.5cm]{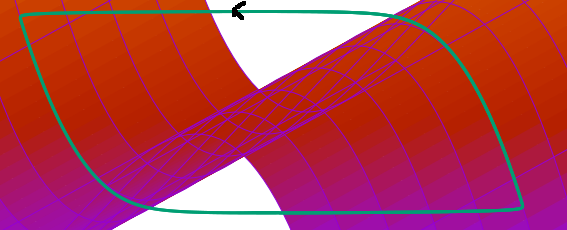}\\[5mm] 
		
			\includegraphics[width=5.5cm,height=4.5cm]{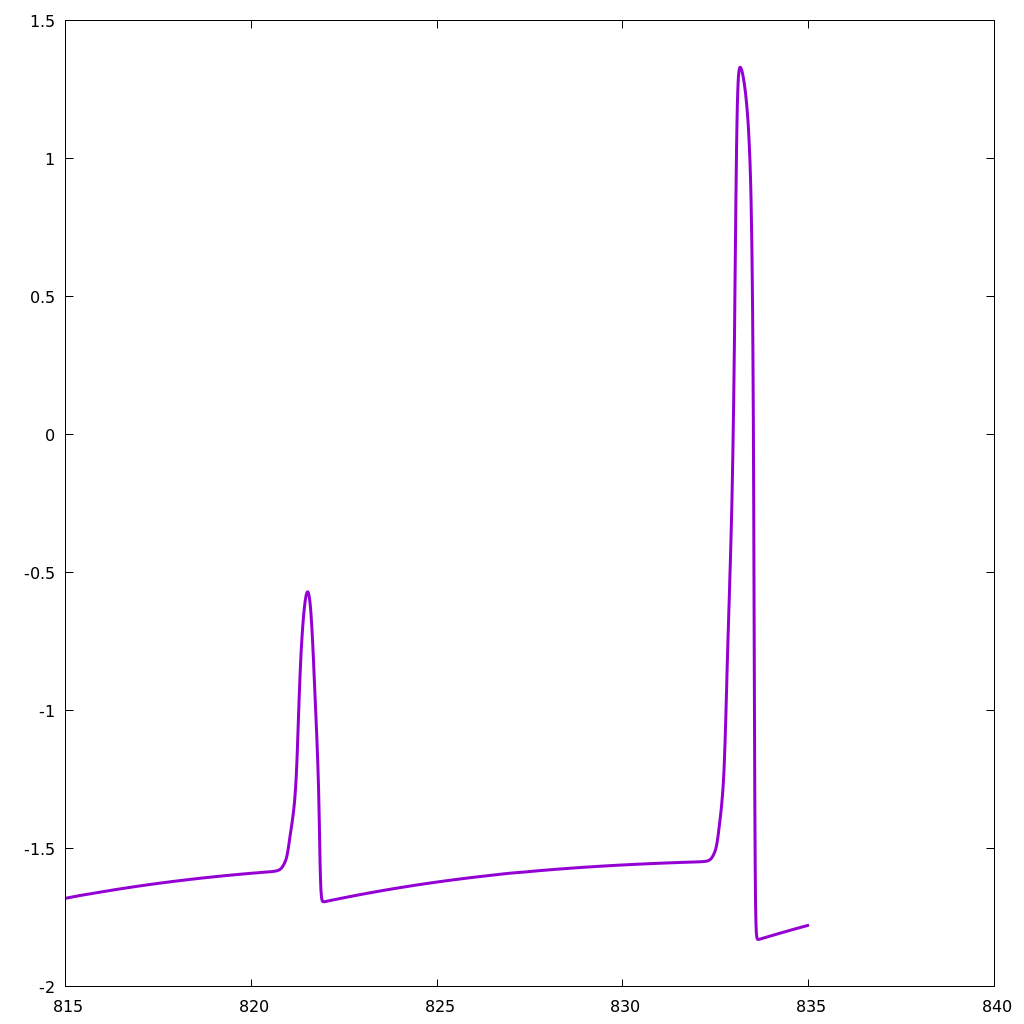}
			\includegraphics[width=5.5cm,height=4.5cm]{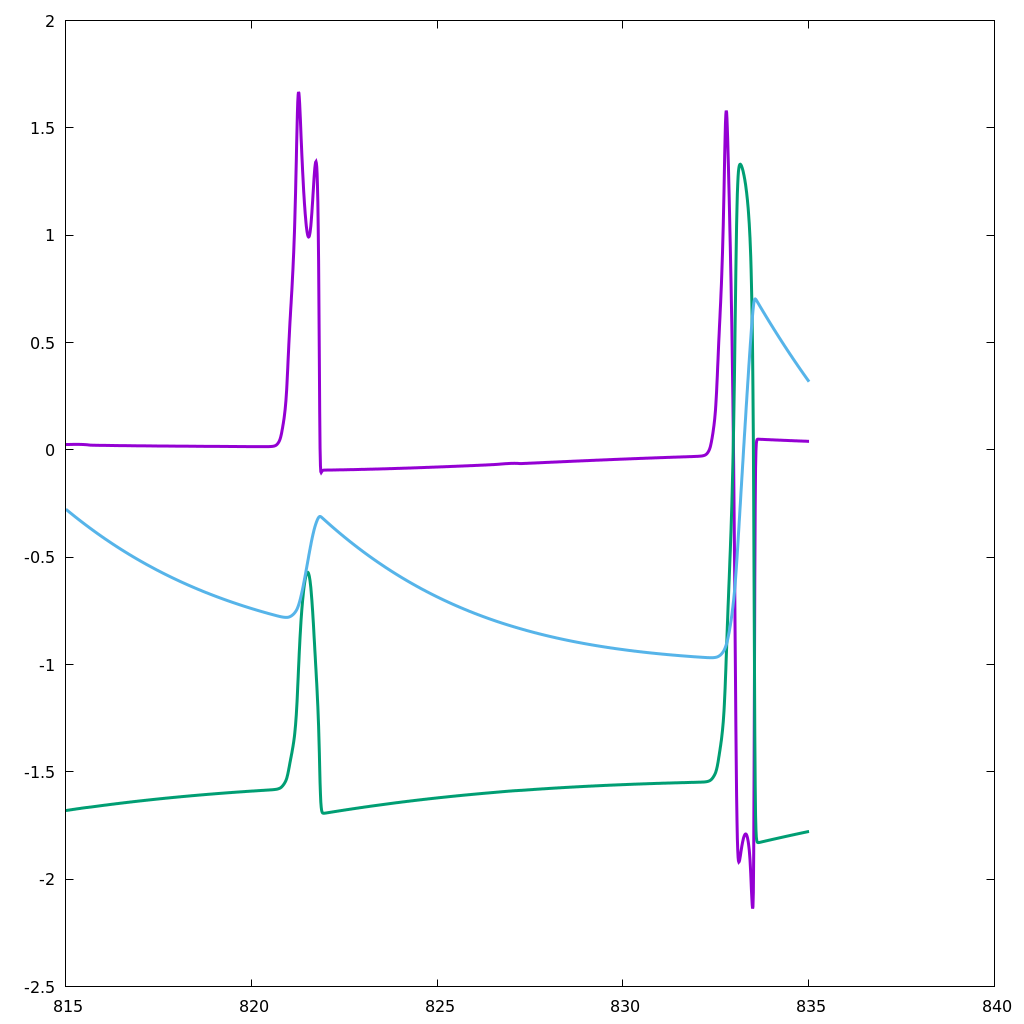}
	\includegraphics[width=5.5cm,height=4.5cm]{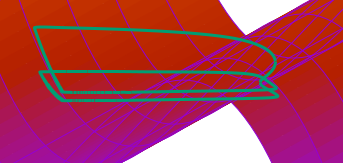}\\[5mm]
	\includegraphics[width=4cm,height=4cm]{FigMMO-PDE2-ut.png}
	\includegraphics[width=4cm,height=4cm]{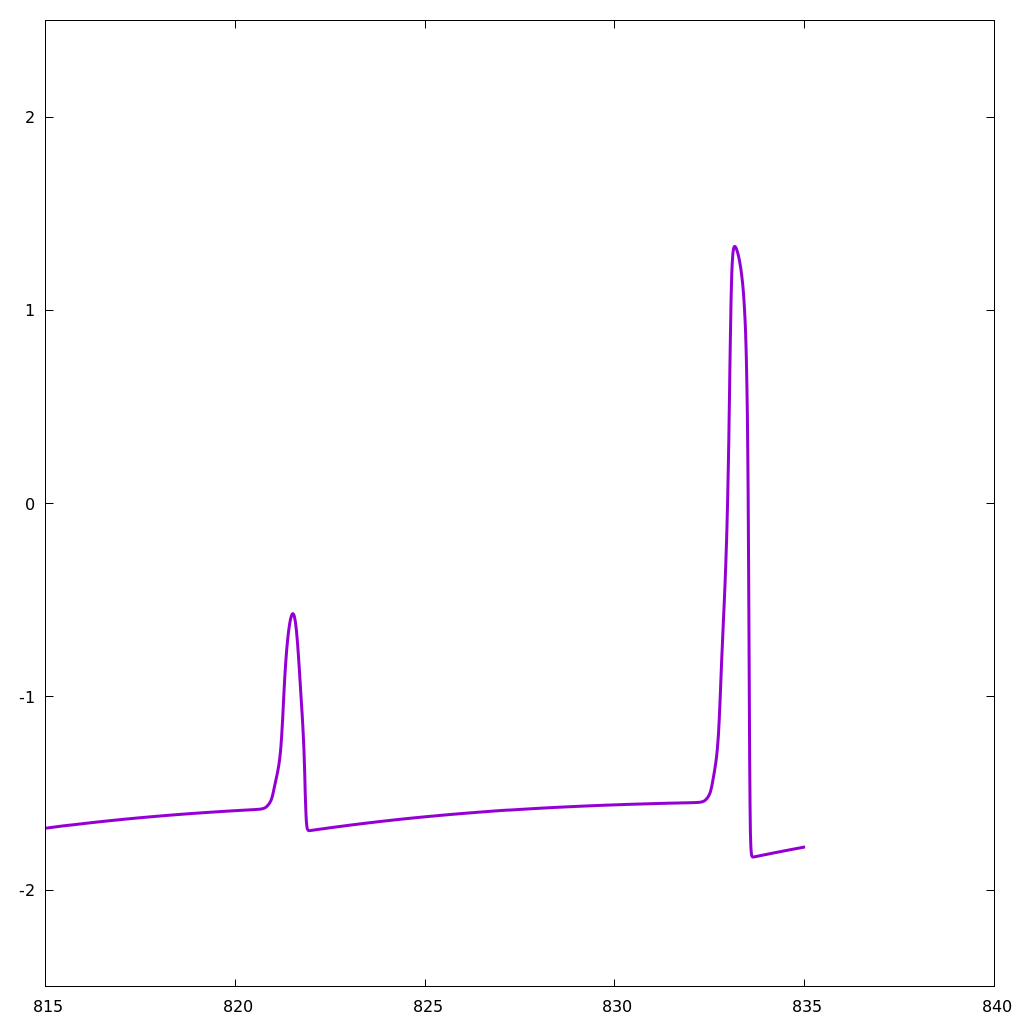}
	\includegraphics[width=4cm,height=4cm]{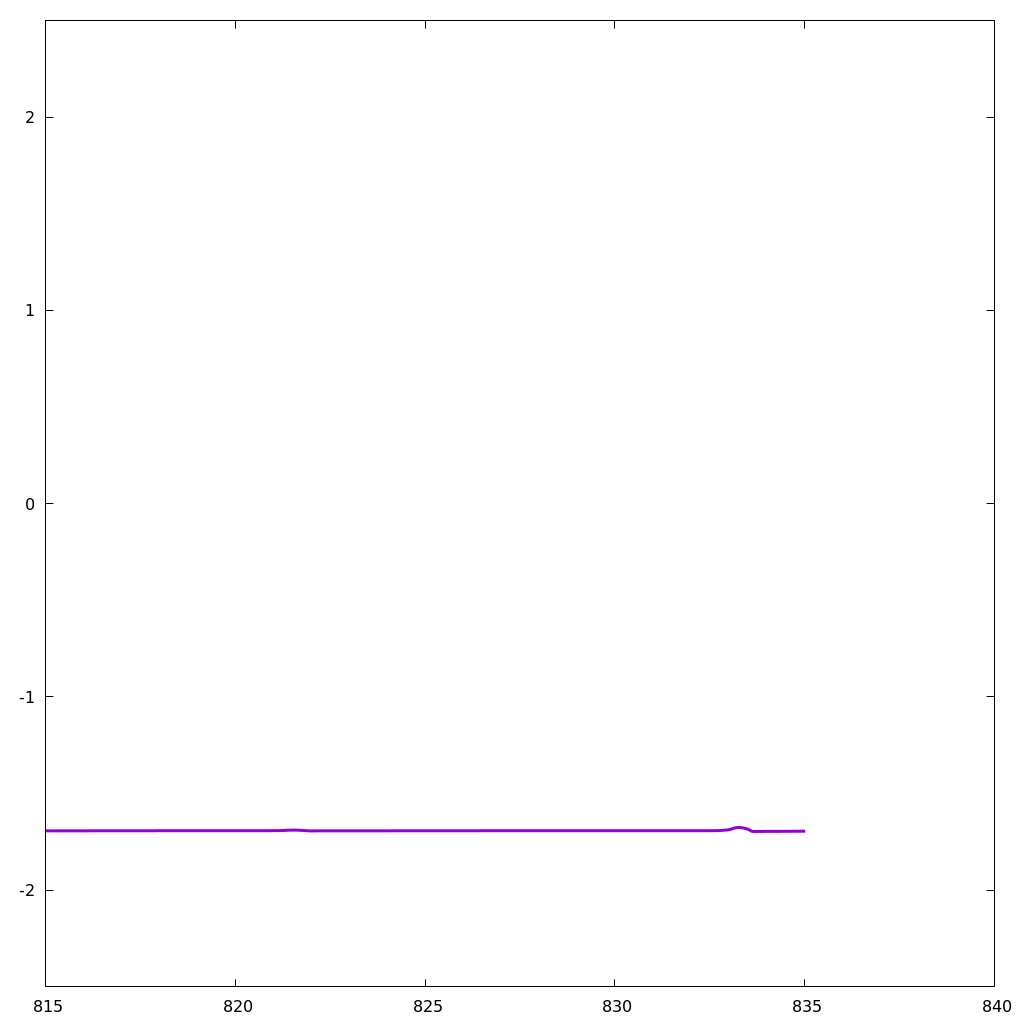}
    \includegraphics[width=4cm,height=4cm]{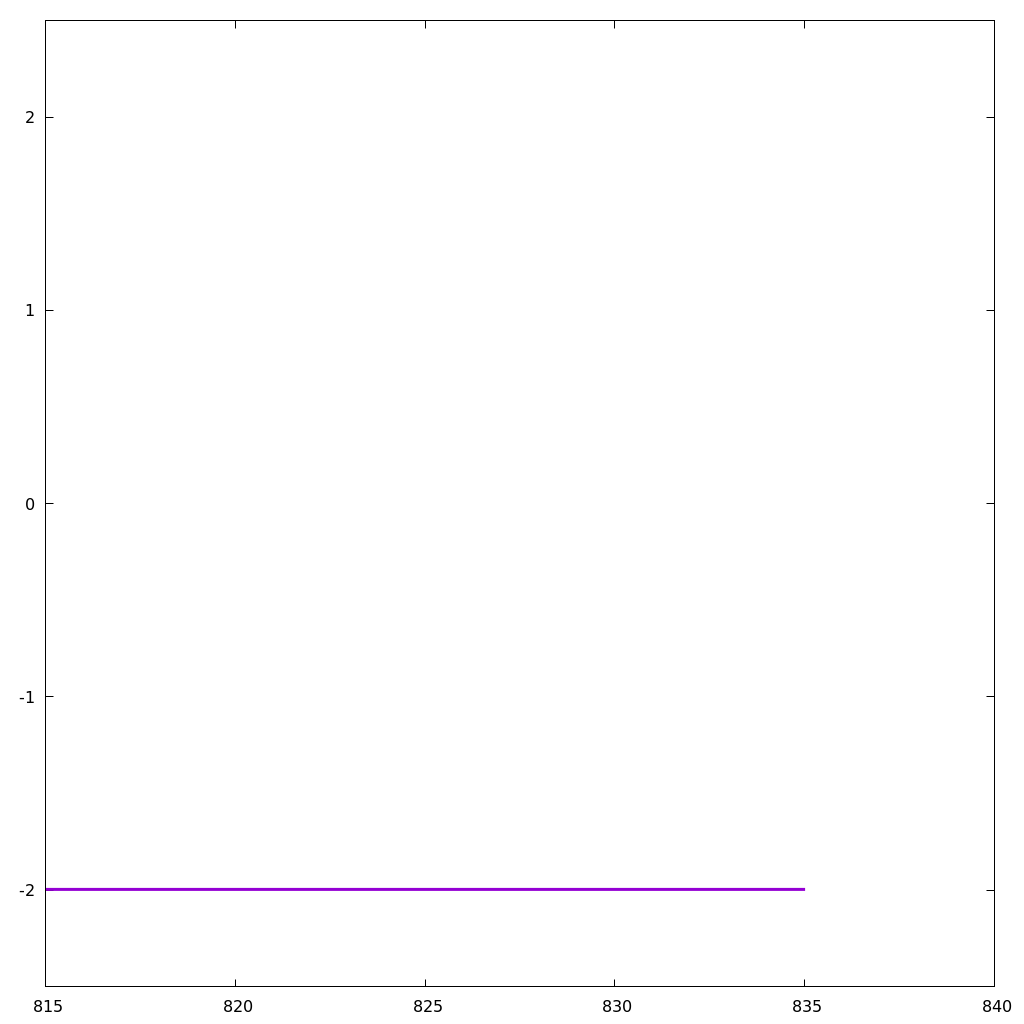}
	
	\caption{This figures illustrates the simulation of \Cref{eq:NhFHN} with the set of parameters given in \ref{eq:parametersPDE2}. The first row corresponds to a fixed value of $x$ near the center for which $c(x)=0$. The second row of the figure corresponds to value of $x$ at an intermediate location between the left border and the center. For these two columns, in the first column, we represent $u$ as a function of time. In the second column, we represent $u,v$ and $u_{xx}$ as functions of time respectively in green, blue and violet. In the third column, we represent the trajectory $u,v,u_{xx}$ along with the manifold $v=f(u)+du_{xx}$ (where $u_xx$ is seen as the third variable). In the last row, we represent $u$ as a function of time. From left to right, we represent four different space locations with the same range of amplitude: left panel corresponds to a center cell, the far right corresponds to a cell close to the left border, the two other panels correspond to intermediate cells. This illustrates the fade of the wave propagation. We refer to \Cref{parag:MMOSPDE2} for more detailed explanation.    }
	\label{fig:PDE2}
\end{figure}

 In this case, again, as in the previous paragraph, if $p$ is close enough to $-1$, waves propagate from the center at the frequency of the diffusionless system (with $c=0$). On the other hand, if $p$ is decreased enough, the system generally converges to a stationary solution. In between, for some range of $p$, regular waves propagate from center but fail to propagate at some point in space. The existence of Hopf Bifurcation has been proved for this specific case, see \cite{Amb-2017}. Related failure phenomenon has been described previously for example in\cite{Yan-2005}. For relevance in biological context, we also refer to \cite{Mai-2013}. Related phenomenon has also been considered in chains of kicked FHN neurons, see \cite{Amb-2021}. The phenomenon is illustrated here in \Cref{fig:PDE2}. At the center, oscillations are as in the ODE diffusionless system. But at some point the oscillation fails. For some intermediate cells, we observe alternance of medium (to compare to \Cref{fig:3}) and larger oscillations. In this case, every other time, the spike will be shortened-- note the difference in the $u_{xx}$ time series.  In this figure, the first row corresponds to a fixed value of $x$ near the center.  At this point of the space $c(x)=0$. The second row of the figure corresponds to an intermediate value of $x$ between the center and the left border. For the two first rows, in the left panel, we represent $u$ for fixed $x$ as a function of time.  In the center panel, we represent $u,v$ and $u_{xx}$ as functions of time, respectively with the colors green, blue and purple.  The right panel illustrates the dynamics of $u,v$ and $u_{xx}$ in the three dimensional space. The ``critical" manifold $v=f(u)+du_{xx}$ is also represented where $u_{xx}$ is considered as a variable. In the last row, we represent $u$ as a function of time. From left to right, we represent four different space locations with the same range of amplitude: left panel corresponds to a center cell, the far right corresponds to a cell close to the left border, the two other panels correspond to intermediate cells. This illustrates the fade of the wave propagation.
 \clearpage
 \clearpage

\bibliography{references}
\section*{Apendix}
\subsection*{A The Topological Degree of Leray-Schauder}
Here, we follow the definition of \cite{Dro-2006}, see also \cite{Bro-2014,Smo-1994,Vol-1994}.
\begin{theorem}
Let $E$ denote a Banach space, and $\mathcal{A}$ the set of $(Id-f,\Omega,y)$ with $\Omega$ an open bounded set of $E$, $y\in E$ and  $f$ a compact map from $\overline{\Omega}$ into $E$ such that $y\notin (Id-f)(\delta\Omega)$. Then, there exists a mapping $d$ from $\mathcal{A}$ to $\Z$ such that
\begin{itemize}
    \item if  $\Omega$ is an open bounded set of $E$ and $y \in E$ then $d(Id,\Omega, y) = 1$.
    \item if  $\Omega$ is an open bounded set of $E$ $y \in E$, $f : \bar{\Omega} \rightarrow E$ is compact and $\Omega_1$, $\Omega_2$ are two open sets in $\Omega$ with $\Omega_1 \cap \Omega_2 =0$ such that  $y \notin (Id-f)(\Omega \backslash (\Omega_1 \cup \Omega_2))$, then 
    \[d(Id-f, \Omega, y) = d(Id-f, \Omega_1, y)+d(Id-f, \Omega_2, y)\]
    \item  if  $\Omega$ is an open bounded set of $E$, $h : [0, 1] \times \Omega \rightarrow E$ is compact,  $y :[0, 1] \rightarrow E$ is continuous and for all $t\in [0,1], y(t) \notin (Id-h(t,\cdot))(\partial \Omega) $ then
    \[d(Id-h(0,\cdot),\Omega,y(0))=d(Id-h(1,\cdot),\Omega,y(1))\]
    $d$ is called the topological degree of Leray-Schauder.
\end{itemize}
\end{theorem}
\begin{prop}
The topological degree of Leray Schauder satisfies
\begin{itemize}
    \item If $d(Id-f,\Omega, y)\neq 0$ there exists $x\in \Omega$ such that $x-f(x)=y$
    \item for all $z \in E$, $d(Id-f,\Omega,y) = d(Id- f- z, \Omega, y-z)$
    \item Let $ (Id-f,\Omega,y) \in \mathcal{A}$ and assume $r=dist(y,(Id-f)(\partial \Omega)) > 0$. If $g : \Omega \rightarrow \R^n$ is compact, and $z\in \R^n$ such that $\sup_{\partial \Omega}(||g-f||)+||y-z||<r$ then
    \[d(Id-f,\Omega,y)=d(Id-g,\Omega,z)\]
    \item $d(Id-f, \Omega, \cdot)$ is constant on the connected components of $E \backslash (d-f)(\partial \Omega)$.
    \item For all $z \in E$, 
    \[d(Id-f,\Omega,y) = d((Id-f)(\cdot,-z),z+\Omega,y).\]
\end{itemize}
\end{prop}

\end{document}